\documentclass[%
 reprint,
 amsmath,amssymb,
 aps,
]{revtex4-2}

\usepackage[dvipsnames]{xcolor}
\usepackage[margin=1in]{geometry} 
\usepackage{amsmath,amsthm,amssymb,amsfonts, mathtools, tensor, siunitx}
\usepackage{hyperref}
\usepackage{braket}

\hypersetup{
     colorlinks   = true,
     citecolor    = magenta,
     urlcolor     = blue,
     linkcolor    = magenta
}

\usepackage{graphicx}
\usepackage{braket}

\usepackage[normalem]{ulem}

\begin{document}
 
 
\title{Three-flavor, Full Momentum Space Neutrino Spin Oscillations in Neutron Star Mergers} 

\author{Henry R. Purcell}
\email{henrypurcell@berkeley.edu}
\affiliation{Department of Physics, University of California, Berkeley, CA 94720}

\author{Sherwood Richers}
\email{richers@utk.edu}
\affiliation{Department of Physics, University of California, Berkeley, CA 94720}
\affiliation{Department of Physics and Astronomy, University of Tennessee, Knoxville, TN 37996}

\author{Amol V. Patwardhan}
\email{apatward@umn.edu}
\affiliation{SLAC National Accelerator Laboratory, Menlo Park, CA 94025}
\affiliation{School of Physics and Astronomy, University of Minnesota, Minneapolis, MN 55455}

\author{Francois Foucart}
\email{francois.foucart@unh.edu}
\affiliation{Department of Physics \& Astronomy, University of New Hampshire, 9 Library Way, Durham NH 03824, USA}

\begin{abstract}
    In the presence of  anisotropic neutrino and antineutrino fluxes, the quantum kinetic equations drive coherent oscillations in neutrino helicity, frequently referred to as spin oscillations. These oscillations depend directly on the absolute mass scale and Majorana phase, but are usually too transient to produce important effects. In this paper we present a full momentum-space analysis of Majorana neutrino spin oscillations in a snapshot of a three-dimensional neutron star merger simulation. We find an interesting angular dependence that allows for that resonant and adiabatic oscillations to occur along specific directions in a large volume of the merger remnant. The solid angle spanned by these directions is extremely narrow in general. We then analyze spin transformation in the presence of flavor transformation by characterizing how the effect's resonance and timescale change during a fast flavor instability. For this analysis, we derive a generalized resonance condition that poses a restrictive requirement for resonance to exist in any flavor channel. We determine that spin oscillations at all locations in the merger snapshot have a length scale that is too large for significant oscillations to be expected even where there exist resonant and adiabatic directions.

\end{abstract}
 
\maketitle





\section{Introduction}

Neutrino transport plays a crucial role in the dynamical evolution of neutron star mergers (NSMs) and core collapse supernovae (CCSNe). In neutron star mergers, abundant weak nuclear interactions give rise to a massive outward neutrino flux that crucially influences the mass outflow and nucleosynthesis evolution of the event \cite{radice_review_2020, Wanajo2014, merger_simulation, Radice:2021jtw, Foucart:2022kon, Cusinato_2022}. Importantly, the relative strengths of the electron neutrino and antineutrino absorption and emission rates determine the equilibrium ratio of protons to neutrons and the consequent fate of nucleosynthesis, thereby governing the proportions of elements that are formed in the ejecta (e.g., \cite{qian1993connection}). Neutrinos are similarly important in canonical CCSNe, and in addition are the factor that fundamentally drives the explosion itself (e.g., \cite{janka_ExplosionMechanismsCoreCollapse_2012, mezzacappa_RealisticModelsCore_2022}).

As a consequence, our understanding of NSMs and CCSNe is only as good as our comprehension of the complex neutrino processes that transpire in these events. A large body of research has focused on neutrino transport effects (e.g., \cite{foucart_NeutrinoTransportGeneral_2023,mezzacappa_PhysicalNumericalComputational_2020a}), as well as exploring methods for modeling flavor transformation and probing sensitivity to neutrino flavor transformation effects (e.g., \cite{Duan2010, Chakraborty+2016, Tamborra2020, Richers_Sen_2022, capozzi_review_2022} and references therein). 

In addition to neutrino transport and flavor transformations, another phenomenon within the neutrino sector that could emerge in NSMs and CCSNe involves neutrinos potentially undergoing helicity or spin transformation, which would drive active-to-sterile conversions for Dirac neutrinos and neutrino-antineutrino conversion for Majorana neutrinos (see e.g., \cite{Volpe:2016xxd, Dobrynina:2016rwy, Studenikin:2016oyh} and references therein). Although such conversions have yet to be demonstrated experimentally, a significant amount of helicity transformation would change the rate of energy transport and the path of nucleosynthesis in NSMs and CCSNe. If such an astrophysical site exists, inputs from probes of the fundamental Dirac or Majorana nature of the neutrino \cite{Elliott_2015_majorana}, the existence of sterile neutrinos \cite{dasgupta_SterileNeutrinos_2021}, and the neutrino masses \cite{degouvea_NeutrinoMassModels_2016} will be crucial to developing detailed models.

Helicity oscillations (also referred to in various contexts as spin or spin-flavor oscillations) of Dirac as well as Majorana-type neutrinos  were first proposed as an explanation to the solar neutrino deficit~\cite{Cisneros:1970nq, Okun:1986hi, Okun:1986na, Okun:1986uf, Voloshin:1986ty, Akhmedov:1987nc, Lim:1987tk}, before the Mikheyev-Smirnov-Wolfenstein (MSW) resonance eventually gained acceptance as the favored solution to this problem. At the same time, it was recognized that these effects could also be present in supernovae and neutron star environments~\cite{Fujikawa:1980yx, Schechter:1981hw, Lim:1987tk}, and the prospect of resonantly-enhanced helicity oscillations was also identified~\cite{Lim:1987tk,Akhmedov:1987nc}.

 
Neutrino spin oscillations can be driven in several ways. These early works considered the scenario where external magnetic fields can couple to a nonzero Dirac or Majorana neutrino magnetic dipole moment to induce spin precession, an effect that has since been further explored in more detail~\cite{Studenikin:2020mfh, deGouvea:2012hg, deGouvea:2013zp, Chukhnova:2020xth, Ternov:2016njz, Bulmus:2022gyz, Lobanov:2021hye, Chukhnova:2020pct, Grigoriev:2018cvo, Chukhnova:2019oum, Sasaki:2021bvu, Abbar:2020ggq, Dash:2018cez, Yuan:2021exm, Jana:2022tsa, Kharlanov:2020cti, Sasaki:2023sza, Giunti:2020dqw}. This can also serve to probe the Dirac/Majorana nature of the neutrino \cite{Alok:2022ovy}. In the standard model, a small magnetic moment is tied to the neutrino mass, but in beyond-standard-model physics there can be additional contributions to the neutrino magnetic moment that are unrelated to the mass (e.g., \cite{Cisneros:1970nq, Studenikin:2020mfh}). Previous research has shown that this effect is only significant in the presence of extremely large magnetic fields, large neutrino moments (typically larger than the standard model value based on limits on neutrino masses), or both, although resonances can arise that enhance helicity transformation probabilities (e.g., \cite{Lim:1987tk, Akhmedov:1987nc, Chukhnova:2019oum, Chukhnova:2020xth, Sasaki:2023sza}). Alternatively, spin-flip effects can be driven by strong gravitational fields or alternate theories of gravity \cite{2021PhRvD.104b4021M,P_riz_1996, Sorge_2007, DVORNIKOV_2006, PhysRevD.101.024016, PhysRevD.55.7960, Chakraborty_2015, Ahluwalia_1996, Lambiase_2005}, which more generally can drive helicity conversions for any spinning particle.
 
Finally, spin oscillations can be driven by an anisotropic background due to an axial-vector potential generated by neutrino forward scattering on polarized or asymmetric matter or neutrino currents \cite{NQK,newspin,Vlasenko:2014bva,Volpe:2013uxl,serreau_volpe_2014}. 
These anisotropy-induced spin oscillations, which are the focus of this paper, occur independently of an intrinsic magnetic dipole moment and can also affect both Dirac and Majorana Neutrinos. This form of helicity oscillations imposes a dependence of the QKEs on the neutrino absolute mass scale and Majorana phases, potentially leading to observable manifestations of these parameters.

Recent studies developed the formalism for extending the neutrino evolution equations to include helicity coherence induced by anisotropic matter/neutrino backgrounds~\cite{Volpe:2013uxl, Vaananen:2013qja, serreau_volpe_2014, NQK, newspin, Kartavtsev:2015eva, Dobrynina:2016rwy}, as well as neutrino-antineutrino pair correlations~\cite{Volpe:2013uxl, Vaananen:2013qja, serreau_volpe_2014, Kartavtsev:2015eva}. The magnitude of this helicity-changing Hamiltonian is proportional to the neutrino mass and inversely proportional to the neutrino momentum, leading to helicity oscillations with negligible amplitudes in most circumstances. It was recognized, however, that helicity transformations through this channel could also be resonantly enhanced~\cite{newspin,Kartavtsev:2015eva}. Significant helicity transformations require not only the presence of a resonance, but the resonance must also be adiabatic so that the oscillations occur on a sufficiently short timescale compared to the evolution of the background (i.e., before the resonance is lost). It was demonstrated subsequently that a potential nonlinear amplification mechanism could enhance the prospects of resonant helicity transformation for Majorana neutrinos~\cite{Vlasenko:2014bva}, using a simple parametric form for the neutrino forward-scattering potential from ordinary matter. 

Later, Refs.~\cite{Tian2017,Chatelain:2016xva} performed numerical investigations of this effect in CCSN/NSM settings. Ref.~\cite{Tian2017} investigated spin-coherence prospects for Majorana neutrinos during the deleptonization burst phase of a CCSN using both 2-flavor and 3-flavor treatments. It was found that the nonlinear amplification mechanism from~\cite{Vlasenko:2014bva} is not able to operate effectively unless experimentally forbidden values of neutrino mass ($\sim 10$\,eV) were used. Nevertheless, it was noted that, \textit{if} spin transformations were to take hold during this phase of the explosion, it would have substantial implications for the ensuing flavor evolution and the expected neutrino signal at detectors.  Ref.~\cite{Chatelain:2016xva} likewise explored this phenomenon in an NSM environment for both Dirac and Majorana neutrinos, using a 2-flavor calculation, and found no significant spin transformations, even when the neutrino mass was artificially increased to $100$\,eV.

In both works, spin oscillations were found to be too transient to produce significant transformations. However, these models contained many simplifications. For example, both studies invoked the \textit{single-angle approximation} 
(albeit in different geometries) wherein background neutrinos travelling along different trajectories are all assumed to follow an identical flavor (and helicity) evolution history as the test neutrino. Ref.~\cite{Tian2017} assumed a spherically symmetric neutrino \lq bulb\rq\ source to model neutrino emission from a CCSN, whereas Ref.~\cite{Chatelain:2016xva} modeled the neutrino source as a disk, to represent the central object formed in the immediate aftermath of an NSM. The imposition of the single-angle approximation explicitly forbids the fast flavor instability~\cite{Tamborra2020,Richers_Sen_2022}. In \cite{Tian2017}, this permitted a physical separation of the respective regimes where helicity and flavor conversion occur. In particular, the helicity resonance was found to \textit{precede} flavor conversion, and hence only one spin-resonance (the $\nu_e$-$\bar\nu_e$ channel) needed to be considered. Ref.~\cite{Chatelain:2016xva} examined the interplay between the spin resonance and flavor evolution induced by a Matter-Neutrino Resonance (MNR)~\cite{Malkus:2012ts, Malkus:2014iqa, Malkus:2015mda, Wu:2015fga, Vaananen:2015hfa, Zhu:2016mwa, Frensel:2016fge, Tian:2017xbr, Shalgar:2017pzd, Vlasenko:2018irq}, but no significant spin transformations were observed, even when resonances in multiple channels (e.g., $\nu_e$-$\bar\nu_e$, $\nu_e$-$\bar\nu_x$, etc.) were considered.

To add to this complexity, if realized in NSMs, the fast flavor instability (FFI) has the potential to significantly transform neutrino distributions due to coherent neutrino-neutrino interactions. It has been shown that so-called crossings in the angular neutrino distribution yield instability \cite{ELNcrossing,dasgupta_2022_crossing}, and that the FFI is largely ubiquitous in neutron star mergers \cite{wu_2017_ubiquitous,li_siegel_2021,just_2022_merger_flavor}. The interplay between flavor transformation and collisions (e.g.,\cite{Richers_2019_NeutrinoQuantumKinetics,johns_2023_collisional,kato_2023_emission_absorption}) and the net effect of flavor transformation on global scales (e.g., \cite{nagakura2023global}) are very active areas of research, but the FFI itself appears to be a dominant and lasting effect that significantly changes neutrino flavor distributions. Because of the large and ubiquitous impact of fast flavor transformation, we take a first stab at estimating the impact this instability could have on helicity transformation.

In this paper, we generalize previous analyses of anisotropy-induced spin oscillations by studying the effect in the 3-flavor regime and by considering its full 3-dimensional structure in an anisotropic background and in the context of fast flavor instability in a neutron star merger remnant. In Section~\ref{sec:background} we review the theory behind neutrino flavor and helicity coherence. In Section~\ref{sec:methods} we describe our approach to probing for resonance and adiabaticity of helicity transformations in multidimensional, three-flavor dynamical simulations of neutron star mergers, including our approach to testing the potential impact of flavor instabilities. We present the results of our calculations in Section~\ref{sec:results} and provide concluding remarks in Section~\ref{sec:conclusions}.

\section{BACKGROUND}
\label{sec:background}

In this section we outline the theory behind neutrino spin transformation. In section~\ref{sec:distribution_function} we describe the neutrino distribution function and neutrino flux density. In section~\ref{sec:qke} we describe a generalized, spin-dependent form of the QKEs for a neutrino ensemble, derived in \cite{NQK,volpe2015}, which contain a term capable of generating coherent helicity transformation in anisotropic media. We analytically express the time-evolution operator from the QKEs in terms of a chiral 4-potential $\Sigma^\alpha$ that depends on the neutrino flux density and background matter distribution. 

Throughout this paper we will use natural units $\hbar = c = 1$. We will employ the convention that a tensor $T$ presenting a spacetime index $\alpha$, flavor indices $a,b$, and helicity indices $h,h'$ will be denoted as $[T^\alpha_{hh'}]^{ab}$, unless flavor or helicity indices are left implicit. A spacetime index will always be written explicitly. A tensor with no spacetime index will simply be denoted $T^{ab}_{hh'}.$

\subsection{The Neutrino Distribution function} 
\label{sec:distribution_function}
A weakly interacting neutrino ensemble can be fully described by its set of two-point Green's functions, encoding single-particle propagation amplitudes. Although multi-particle correlations may have important consequences (e.g., Refs.~\cite{Patwardhan:2022mxg, Balantekin:2023qvm, Volpe:2023met} 
and references therein), we follow \cite{NQK,volpe2015} and neglect that possibility in this work. Given annihilation operators $a_{i,\vec p, h}$ ($b_{i,\vec p, h}$) for (anti)neutrinos of mass state $i$, three-momentum $\vec p$, and helicity $h$, we can express these two-point functions as
 \begin{equation}
   \left\langle a_{i,\vec p, h},a^\dagger_{j,\vec p', h'}\right\rangle=(2\pi)^32n_{ij}(\vec p)\delta^{(3)}(p-p')f^{ij}_{hh'}(\vec p),   
\end{equation}
\begin{equation}
\left\langle b_{i,\vec p, h},b^\dagger_{j,\vec p', h'}\right\rangle=(2\pi)^32n_{ij}(\vec p)\delta^{(3)}(p-p')\bar {f}^{ij}_{hh'}(\vec p).
\label{eq:amplitudes}
\end{equation}
 Here the $\left\langle \right\rangle$ brackets denote the quantum mechanical and statistical ensemble average and $n_{ij}=2\omega_i\omega_j/(\omega_i+\omega_j)$ is the normalization factor, where $\omega_i(\vec p)=\sqrt{\vec {p}^{\,2}+m_i^2}$. Note that the correlations 
 $\left\langle a^\dagger ,b^\dagger \right\rangle$ and $\langle a,b \rangle$ pairing neutrinos and antineutrinos should in general also be considered~\cite{serreau_volpe_2014}. However these correlations generate transformation only for very low-energy neutrinos, with DeBroglie wavelength comparable to the length scale of the surrounding astrophysical environment. \cite{serreau_volpe_2014,newspin,PhysRevD.71.093004}. 

The numbers $f^{ij}_{hh'}(\vec p, x)$ and $\bar f^{ij}_{hh'}(\vec p, x)$ (which in general are position dependent) encode the full information of the neutrino ensemble, and can be assembled into distribution functions 
\begin{equation}
\begin{aligned}
    F(\vec p,x)&=\begin{pmatrix}
    f_{LL}&f_{LR} \\ 
    f_{RL}&f_{RR}\\
    \end{pmatrix}\\
    \bar{F}(\vec p,x)&=\begin{pmatrix}
    \bar{f}_{LL}&\bar{f}_{LR} \\ 
    \bar{f}_{RL}&\bar{f}_{RR}\\ \end{pmatrix}
    \end{aligned}
    \label{eq:F_definition}
\end{equation}
Here flavor indices are suppressed, so each $f_{hh'}$ is a square $n_f \times n_f$ matrix, where $n_f$ is the number of neutrino flavors. Diagonal elements $f^{ii}_{hh}$ represent occupation numbers for states of helicity $h$ and mass-state $i$, and off-diagonal components give the quantum coherence of states with different mass and/or helicity. For Majorana neutrinos, the distribution function is entirely described by $F$ and is related to $\bar{F}$ by $f_{hh'}=\bar{f}_{hh'}^T$, where the transpose acts on flavor indices.

Building distribution functions from two-point correlations in this manner can only be done in the mass eigenstate basis; to switch to the flavor basis we make use of the Pontecorvo-Maki–Nakagawa–Sakata (PMNS) matrix $U$ and convert the distribution functions via $f_\mathrm{(flavor)}=Uf_\mathrm{(mass)}U^\dagger$. All quantities in this paper will be expressed in the flavor basis, and explicit flavor indices will be indicated by $a,b$ in a superscript.

It will be convenient to define a position-dependent net neutrino flux density $J^\alpha$ given by
\begin{equation}
J^\alpha(x)=\int \frac{d^3p}{(2\pi)^3}n^\alpha(\vec p)(f_{LL}(\vec p,x)-\bar f_{RR}(\vec p,x)),
\label{eq:flux}
\end{equation}
where $J^0$ gives the neutrino number density and the spatial components $J^i$ encode the number flux density. $n^\alpha=(1,\hat{p})$ is the neutrino propagation direction. The full net neutrino current density over all flavors is then $\mathrm{Tr}_\mathrm{f} J^\alpha$, where $\mathrm{Tr}_\mathrm{f}$  indicates a trace over flavor indices.

\subsection{Neutrino Spin Oscillations in the QKEs}  
\label{sec:qke}

The neutrino quantum kinetic equations (QKEs) characterize the evolution of the distribution functions $F$ and $\bar{F}$. Here we outline the QKEs for a Majorana neutrino ensemble, including the important generalizations pioneered in \cite{Vlasenko:2014bva,NQK,newspin,Volpe:2013uxl,Vaananen:2013qja,serreau_volpe_2014} 
that lead to terms which generate coherent helicity oscillations. Specifically, we use QKEs that are derived including spin degrees of freedom and second-order contributions in a power counting scheme defined by ratios of relevant quantities (neutrino masses, mass-splitting, forward scattering-induced matter potentials, and external gradients) to the characteristic neutrino energy in the ensemble.

The QKEs are
\begin{equation}
\begin{aligned}
&D_{\vec p, x}F(\vec p,x)=-i[H(\vec p,x),F(\vec p,x)]+C(\vec p,x), \\
&\bar D_{\vec p, x}\bar F(\vec p,x)=-i[\bar H(\vec p,x),\bar F(\vec p,x)]+\bar C(\vec p,x)
\end{aligned}
\end{equation}
where $D_{\vec p,x}$ is a differential operator generalizing the Vlasov term of transport equations, $H(\vec p, x)$ is a Hamiltonian-like coherent evolution operator, and $C(\vec p,x)$ is an inelastic collision term encoding direction-changing interactions. We neglect the collision term in this paper, as the length scale of inelastic interactions is far larger than that of the fast-flavor instability (e.g., \cite{PhysRevLett.122.091101,Richers_2019_NeutrinoQuantumKinetics}). However, we will show that the characteristic helicity transformation length scale is large compared to the collisional length scale, so if helicity transformations are present, the collisional term would be needed for realistic evolution. $\bar H(\vec p, x)$ and $\bar C(\vec p,x)$ are antineutrino analogues of these operators.

Before continuing, we must define a spacetime coordinate basis formed by two lightlike 4-vectors $n^\alpha(\vec p)=( 1,\hat p),$ $\bar n^\alpha(p)=( 1,-\hat p )$ and two transverse 4-vectors $x_{1,2}^{\alpha}=( 0,\hat x_{1,2})$. These satisfy $n\cdot x_i=\bar n\cdot x_i=0,$ $x_i\cdot x_j=\delta_{ij}$, and we demand that $\hat p, \ x_1,$ and $x_2$ form a right-handed triad. The choice of $x_1$ and $x_2$ is then arbitrary up to a rotation around $n,$ but the algebra turns out to be simplest with the `standard gauge' defined by taking $x_1$ to have the same azimuthal angle  as $n^\alpha$ \cite{newspin, NQK}. With these conditions, the basis is fully defined by the direction of $\vec p$, and can be written in spherical coordinates as
\begin{equation}
\begin{aligned}
&n^\alpha=(1, \ \mathrm{cos}(\phi)\mathrm{sin}(\theta), \  \mathrm{sin}(\phi)\mathrm{sin}(\theta), \cos(\theta)) \\
&x^\alpha_1=(0, \ \mathrm{cos}(\phi)\mathrm{cos}(\theta), \ \mathrm{sin}(\phi)\mathrm{cos}(\theta), \ -\sin(\theta)),\\
&x^\alpha_2=(0, \ -\mathrm{sin}(\phi), \ \mathrm{cos}(\phi), \ 0),
\end{aligned}
\end{equation}
where $\theta$, $\phi$ are the polar and azimuthal angles of $\vec p$ (the $\theta=0$ and $\phi=0$ directions are in reference to an arbitrary choice of orthonormal coordinates).

The coherent evolution operator $H(\vec p, x)$ can be broken up into chiral components as
\begin{equation}
\begin{aligned}
H(\vec p,x)=\begin{pmatrix}H_{L}&H_{LR} \\ H_{LR}^\dagger&H_{R} \end{pmatrix},\\
\bar H(\vec p,x)=\begin{pmatrix}\bar H_{L}&H_{LR} \\ H_{LR}^\dagger&\bar H_{R} \end{pmatrix},
\end{aligned}
\label{eq:H_chiral}
\end{equation} 
where each element on the RHS of Equation~\ref{eq:H_chiral} is a $3\times3$ matrix (flavor indices are omitted). The off-diagonal block $H_{LR}$ drives coherent oscillations between left- and right-handed neutrinos, while the diagonal blocks $H_L$ and $H_R$ drive neutrino flavor change. We can express these as \cite{NQK,newspin}
\begin{equation}
\begin{aligned}
&H_R=\Sigma^\kappa_R+\frac{1}{2|\vec p|}(m^\dagger m-\epsilon^{ij}\partial^i\Sigma ^j_R+4\Sigma^+_R\Sigma^-_R)  
\\  
&H_L=\Sigma^\kappa_L+\frac{1}{2|\vec p|}(m m^\dagger-\epsilon^{ij}\partial^i\Sigma ^j_L+4\Sigma^-_L\Sigma^+_L) 
\\
&H_{LR}=-\frac{1}{|\vec p|}(\Sigma_R^+m^\dagger-m^\dagger\Sigma_L^+).   
 \label{eq:HR}
\end{aligned}
\end{equation}
The $\Sigma^\kappa_{L,R}$ and $m^\dagger m/2|\vec p|$ terms in the diagonal components of $H_L(\vec p, x)$ and $H_R(\vec p, x)$ are present in standard analyses of neutrino oscillations. The first encodes forward-scattering on background matter and other neutrinos and the second controls vacuum oscillations. All other terms, including the off-diagonal spin-flip component $H_{LR}$, arise only in the presence of helicity coherence \cite{NQK,newspin,volpe2015}. Here $m$ is the neutrino mass matrix, for which we describe our particular assumptions in Section~\ref{sec:methods}. Since in this paper we ignore inelastic collision terms and focus on state dynamics for a given $\vec p$ and $x$, we  refer to $H(\vec p, x)$ simply as the Hamiltonian of the system.

$\Sigma^\kappa_{L}$ ($\Sigma^\kappa_{R}$), $\Sigma^i_{L}$ ($\Sigma^i_{R}$), and $\Sigma^\pm_{L}$ ($\Sigma^\pm_{R}$) are projections onto the previously mentioned basis of a chiral 4-potential $\Sigma^\alpha$ generated by forward scattering on background matter and neutrinos:
\begin{equation}
    \Sigma^\alpha(x)=\begin{pmatrix}\Sigma^\alpha_L(x)&0\\0& \Sigma^\alpha_R(x)\end{pmatrix}.
\end{equation}
Specifically, $\Sigma^\kappa_{L,R}=n_\alpha(\vec p)\Sigma_{L,R}^\alpha$ and $\Sigma^i_{L,R}=x_i\cdot\Sigma_{L,R}^\alpha$, and $\Sigma^\pm_{L,R}\equiv (1/2)e^{\pm i\phi}(x_1\pm i x_2)_\alpha\Sigma^\alpha_{L,R}$.

\begin{figure}
    \includegraphics[width=\linewidth]{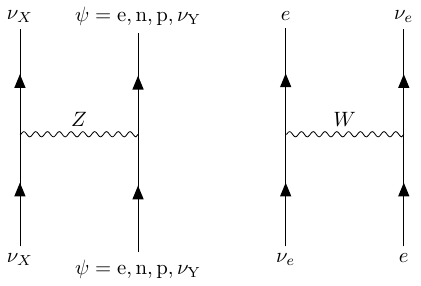}
 \caption{ Tree-level Feynman diagrams encoding forward scattering contributions to $\Sigma^\alpha$ \cite{NQK}}
 \label{fig:diagrams}
 \end{figure}

$\Sigma^\alpha$ is generated by forward scattering on electrons, nucleons, and background neutrinos, specifically receiving contributions from the Feynman diagrams in Figure \ref{fig:diagrams}. The total potential factoring into Equation \ref{eq:HR} is the sum of the matter and neutrino contributions:
\begin{equation}
 \Sigma^{\alpha}_R = \left.\Sigma^{\alpha}_R \right|_\nu+\left.\Sigma^{\alpha}_R \right|_\mathrm{mat}. 
 \label{eq:Sigma_tot}
\end{equation}

The potential contributed by background neutrinos is given by
\begin{equation}
\begin{aligned}
\left.[\Sigma_R^{\alpha}]^{ab}\right|_\nu=\sqrt{2}G_F\left([J^{\alpha}]^{ab} +\delta^{ab}\mathrm{Tr}_\mathrm{f}[J^{\alpha}]\right)
\label{eq:Sigma_neutrinos}
\end{aligned}
\end{equation}
where the indices $a$ and $b$ indicate flavor and $[J^{\alpha}]^{ab}$ is the neutrino flux density. The expression for the matter contribution is similar, but can be simplified by assuming an unpolarized, charge-neutral matter background with equal amounts of left- and right-handed electrons and nucleons. This is a good approximation in the absence of strong magnetic fields. Additionally, we transform all quantities to a comoving reference frame such that the local velocity of matter (i.e., everything but neutrinos) is zero and assume that there is not sufficient energy to produce significant numbers of $\mu$ and $\tau$ leptons \cite{radice_review_2020}. These assumptions lead to a matter contribution given by
\begin{equation}
\begin{aligned}
\left.[\Sigma_R^{\alpha}]^{ab}\right|_{\mathrm{mat}} &= \sqrt{2}G_F n_B\left[Y_e\delta^{ea}\delta^{eb} + \frac{1-Y_e}{2}\delta^{ab}\right] u^\alpha
\end{aligned}
\label{eq:sigma_matter}
\end{equation}
where $Y_e$ is the local electron fraction, $n_B$ is the baryon number density, and $u^\alpha=\{1,0,0,0\}$ is the fluid four-velocity. 

The expressions for the antineutrino evolution operator $\bar H$ are identical to those in Equation~\ref{eq:HR}, except that the sign of the terms multiplied by $\frac{1}{2|\vec p|}$ in $H_R$ and $H_L$ are flipped. For Majorana neutrinos, $\Sigma_R=-\Sigma_L^T$, whereas for Dirac neutrinos $\Sigma^\alpha_L\approx 0$ \cite{newspin}, though we limit the present work to Majorana neutrinos for simplicity.

\section{Methods}
\label{sec:methods}
This section details the theoretical and computational techniques employed in our 3-dimensional analysis of spin oscillations and the fast-flavor instability. Part \ref{sec:resonance} outlines the equations we use to identify resonant and adiabatic helicity oscillations.
Part \ref{sec:methodsmerger} describes the neutron star merger simulation \cite{merger_simulation} we used for our analysis, and explains how we reconstructed the time evolution operator of Equation \ref{eq:HR} and relevant spin oscillation parameters from the data of the merger. Part \ref{sec:methods_ffi} describes our particle-in-cell simulation of the QKEs with which we locally simulate the fast flavor instability in the neutron star merger.

Throughout this paper we will use a Normal Hierarchy neutrino mass matrix constructed from the best-fit mass-squared differences and mixing angles given in \cite{Esteban_2020}; specifically we take $\Delta m^2_{21} = 7.42\times 10^{-5} \ \mathrm{eV}^{2}$, $\Delta m^2_{31} = 2.514\times 10^{-3} \ \mathrm{eV}^{2}$, $\theta_{12} = 33.44^\circ,$ $\theta_{13} = 8.57^\circ,$ $\theta_{23} = 49.0^\circ,$ and $\delta_{CP} = 195 ^\circ$. Following \cite{Loureiro:2018pdz} we take $\Sigma m_\nu = 0.26$ eV, a high upper bound to give the best possible chance to spin-flip effects which grow linearly with the neutrino mass. These values lead to masses $m_1 = 0.08227$ eV, $m_2 = 0.08268$ eV, and $m_3 = 0.09505$ eV. We take the magnitude of the neutrino momentum to be $|\vec p | = 10$ MeV, a representative value for NSM neutrinos \cite{foucart_NeutrinoTransportGeneral_2023}. Note that this does not match the location- and species-dependent average neutrino energy, but is sufficient for determining the importance of the helicity-transforming parts of the Hamiltonian and allows for easy rescaling under different assumptions of mass or energy.

\subsection{Conditions for Significant Helicity Oscillations}
\label{sec:resonance}
Similar to the MSW mechanism \cite{mikheyev_smirnov_1985,wolfenstein_1978}, a neutrino can experience significant helicity transformation when it passes through a resonance adiabatically. Reference \cite{Tian2017} quantifies this under the assumption that there are equal fluxes of all heavy lepton neutrino flavors, but in the presence of the FFI this is not necessarily guaranteed. In this section we generalize the criteria for significant helicity transformation so it can be applied in realistic neutron star merger scenarios in the presence of dynamical flavor transformation.

\subsubsection{Resonance}

For significant transformation, the Hamiltonian must be misaligned from the particle's quantum state in a way that sources helicity coherence. We can determine this by looking at the Hamiltonian's eigenvectors. Specifically, assuming a constant Hamiltonian $H$, an initial 6-dimensional (3-flavor, 2-spin) neutrino state vector $\ket{\nu(0)}$ evolves as
\begin{equation}
    \ket{\nu(t)} = \sum_{k{ = 1}}^6a_{k}\ket{\lambda_{k}}e^{-iE_{k}t}
    \label{eq:resonance_derivation_time_evolution}
\end{equation}
where $k$ indexes energy eigenstates and eigenvalues $\ket{\lambda_{k}}$  and $E_{k}$, and $a_{k}$ are the components of $\ket{\nu(0)}$ in the energy basis.
Note that resonance in helicity oscillations cannot occur if the eigenvectors $\ket{\lambda_{k}}$ are all either purely left-handed or purely right-handed; there have to be some eigenvectors that are mixed in helicity with a large component in both the left- and right-handed subspaces. This is because resonance requires that the changing phases $e^{-iE_{k}t}$ cause all the left-handed parts of the eigenvectors to cancel out at some point in time in Equation \ref{eq:resonance_derivation_time_evolution}, and all the right-handed parts to cancel out at a later point in time, so that we get an initially right-handed neutrino transitioning to a left-handed neutrino. 
But eigenvectors are orthogonal, and therefore there is no superposition of purely left-handed (or right-handed) eigenvectors that cancels to 0. We therefore have that a necessary condition for resonance is that at least one eigenvector of the Hamiltonian has 'sufficiently large' left- and right-handed parts.

Quantitatively, for neutrinos with $N_F$ 
flavor states, we show in Appendix~\ref{app:proof} that a necessary condition for helicity resonance is
\begin{equation}
    \Omega\geq 1\,,
    \label{eq:general_resonance}
\end{equation}
where
\begin{equation}
\Omega = 2N_F^2\max_j \left( 1- \left | \vphantom{\sum}  \tensor[_R]{\braket{\lambda_j|\lambda_j}}{_{\!R}}   - \tensor[_L]{\braket{\lambda_j|\lambda_j}}{_{\!L}} \,  \right | \right)\,\,,
\label{eq:Omega}
\end{equation}
$\ket{\lambda_j}$ is the $j$'th eigenvector, and $\ket{\lambda_j}_{L,R}$ is the same eigenvector with all right- and left-handed components respectively set to zero. See Appendix~\ref{app:proof} for a derivation of Equation \ref{eq:general_resonance}. 

This generalized resonance condition allows for a treatment of helicity oscillations applicable to a general mixture of an arbitrary number of neutrino flavors, which was not possible with previous approaches. If a neutrino distribution shows minimal flavor coherence such that the Hamiltonian is approximately flavor-diagonal, we can use arguments analogous to MSW resonance to derive that resonant helicity oscillations are guaranteed between flavor-helicity states $\ket{aL}$ and $\ket{bR}$ if and only if 
\begin{equation}
    |H_L^{aa} - H_R^{bb}| \ll |H_{LR}^{ab}|\,\,.
    \label{eq:resonance_simplified}
\end{equation}
Hence we have a necessary and sufficient condition for helicity resonance between each pair of flavors $a$ and $b$, assuming negligible flavor mixing. In the $e_L\rightleftharpoons e_R$ channel, this gives us the condition $|H_L^{ee} - H_R^{ee}| = 0$ used in \cite{Tian2017}, which can be simplified by inserting the expressions for the Hamiltonian components (equations \ref{eq:HR}, \ref{eq:Sigma_neutrinos}, \ref{eq:sigma_matter}) to obtain
\begin{equation}
    \frac{G_fn_b(3Y_e-1)}{\sqrt{2}}+\left[H^{ee}_L\right]_\nu=0\,\,,
    \label{eq:resonance_simplified_e}
\end{equation}
a necessary and sufficient condition for helicity resonance in electron neutrinos in the flavor-diagonal case.

However, in the context of leptonic interactions, weak magnetism effects, and the multitude of flavor transformation phenomena present in neutron star mergers, flavor mixing will occur and the applicability of Equation~\ref{eq:resonance_simplified_e} will break down. The generalized resonance condition (Equation~\ref{eq:general_resonance}) is more robust in that it applies even in the context of substantial flavor mixing, and encompasses resonance between all possible flavor states in a single equation. This comes at the cost of being a necessary but not sufficient condition, but as we will see in our results it is an extremely strong constraint.

\subsubsection{Adiabaticity}
\label{sec:methods_adiabaticity}

Resonance on its own does not guarantee that significant neutrino spin oscillations will occur. If the neutrino Hamiltonian evolves too non-adiabatically, then the resonance condition may be satisfied so briefly that the resultant spin oscillations are negligible.

If a neutrino is oscillating between a pure left-handed state $\ket{\nu_L}$ and a pure right-handed state $\ket{\nu_R}$, then the angular velocity of helicity transformations is 
\begin{equation}
     \omega_\mathrm{osc} = |\bra{\nu_L}H\ket{\nu_R}|
\end{equation}
to first-order by the Schrodinger equation. Similar to the case of MSW resonance \cite{mikheyev_smirnov_1985}, the angular velocity of the angle between the Hamiltonian and the initial state, which determines how long the state remains in resonance, is
\begin{equation}
    \omega_\mathrm{res} = \frac{\nabla_{\vec p} |\bra{\nu_R}H\ket{\nu_R} - \bra{\nu_L}H\ket{\nu_L}|}{2|\bra{\nu_L}H\ket{\nu_R}|}, 
    \label{eq:resonant_timescale}
\end{equation}
where $\nabla_{\vec p}$ denotes a derivative along the neutrino world line. We can therefore define the \textit{adiabatic index} $\gamma$ by  
\begin{equation}
    \gamma = \frac{\omega_\mathrm{osc}}{\omega_\mathrm{res}} = \frac{2|\bra{\nu_L}H\ket{\nu_R}|^{2}}{ |\bra{\nu_R}\nabla_{\vec p}H\ket{\nu_R} - \bra{\nu_L}\nabla_{\vec p}H\ket{\nu_L}|}\,\,,
    \label{eq:gamma_general}
\end{equation}
and adiabatic oscillations can be expected when 
\begin{equation}
 \gamma \gg 1,
    \label{eq:adiabaticity_general}
\end{equation}
 in which case the time scale of the oscillations is short compared to the period over which the resonance condition is satisfied.
 
 For spin oscillations between left- and right-handed electron neutrinos, this reduces to the expression given in \cite{Tian2017}, namely
 \begin{equation}
\gamma_{ee} = \frac{2 |H_{LR}^{ee}|^2}{\nabla_{\vec{p}}H_{LL}^{ee}}.
\label{eq:adiabaticity_simplified}
\end{equation}
We arrive at this equation by simplifying Equation~\ref{eq:HR}, setting the extremely small latter terms $-\epsilon^{ij}\partial^i\Sigma ^j_R, \ 4\Sigma^+_R\Sigma^-_R \approx 0$ and using the fact that $\Sigma^\kappa_R=-[\Sigma^\kappa_L]^T$ for Majorana neutrinos.  

Note that Equation \ref{eq:resonant_timescale} is a first-order approximation of the timescales over which resonance is preserved. If the Hamiltonian has a very small time derivative along the neutrino trajectory then Equation \ref{eq:resonant_timescale} suggests resonance will be maintained for a long period, but a large second derivative could still mean it will be lost quickly. We discuss this further in our results.

In order to evaluate adiabaticity, we need derivatives of the potential $\Sigma$ defined in a local orthonormal tetrad. Transforming the potential four-vector and the derivative into the coordinate frame yields
\begin{equation}
\begin{aligned}
    \nabla_\alpha \Sigma^\nu &= e^{\widetilde{\alpha}}_{(\alpha)} e^{(\nu)}_{\widetilde{\beta} }\nabla_{\widetilde{\alpha}}\left(e^{\widetilde{\beta}}_{(\delta)}\Sigma^{\delta}\right)\\
    \end{aligned}
\end{equation}
where indices with a tilde indicate coordinate frame quantities, bare indices indicate tetrad frame quantities, and $e$ are the tetrad basis vectors. We evaluate the partial derivatives within the covariant derivative using centered finite diferencing. This is important to account for potential rotations of the tetrad from one location to the next. Note that when we evaluate derivatives on the snapshot, we treat partial time derivatives as zero. This is certainly an approximation, but we choose to do this in lieu of full dynamical quantum kinetic simulations.

Due to the smallness of the $m^\dagger/|\vec p|$ factor in Equation \ref{eq:HR}, 
$H_{LR}$ (and therefore the energy gap 
at resonance and $\gamma$) tends to be very small for supernova or merger neutrinos with characteristic momentum $|\vec p| = 10\,\mathrm{MeV}$. Indeed, \cite{Tian2017} have shown that only the lowest energy neutrinos satisfy $\gamma\gg1$, limiting the feasibility of significant spin oscillations occurring in neutron star mergers. 

One objective of this paper, then, is to determine whether the shifts in neutrino flux caused by the fast flavor instability can increase the magnitude of $H_{LR}$ enough to make helicity oscillations important.

\subsection{Neutron Star Merger Background}
\label{sec:methodsmerger}

The background matter, neutrino, and spacetime background are taken from a simulation snapshot $5\,\mathrm{ms}$ after the merger of two  $1.2M_\odot$ neutron stars\cite{merger_simulation}. The merger simulation is performed with the SpEC code, which evolves Einstein's equations of general relativity, the relativistic fluid dynamics equations. Neutrinos are evolved using a two-moment transport scheme evolving the energy integrated neutrino energy density, momentum flux density, and number density for each species of neutrinos and antineutrinos (with muon and tau neutrinos and antineutrinos assumed to all have the same distribution function). From these quantities, we can also directly calculate for each species the average energy of the neutrinos, which we use to construct the the number four-flux vectors. The dense matter is described by the LS220 equation of state~\cite{Lattimer:1991nc}. We use data only from a single refinement level with grid spacing of $0.68\,\mathrm{km}$ that covers a domain of size $136\times136\times68\,\mathrm{km}$.

The two-flavor neutron star merger simulation developed in \cite{merger_simulation} carries a three dimensional grid of data for the (anti)neutrino number density $n_{\nu_i}$ ($\bar n_{\nu_i}$) and spatial number flux $\widetilde{F}_{\nu_i}(x)$ ($\bar {\widetilde{F}}_{\nu_i}(x)$), where the subscript $i=e, x$ denotes either the electron or heavy-lepton flavor. The neutrino lepton current $J^\alpha(x)$ of Equation \ref{eq:flux} can be expressed in terms of these as 
\begin{equation}
\begin{aligned}    
[J^\alpha]^{aa}(x) &= (n_{\nu_a}-\bar n_{\nu_a}, \widetilde{F}_{\nu_a}(x) - \bar{\widetilde{F}}_{\nu_a}(x)),
\end{aligned}
\end{equation}
and all other flavor components are zero. We transform the currents into an orthonormal tetrad comoving with the background fluid. We set the tetrad's $\hat{z}$-axis to point along net lepton number flux ($\sum_a [J^i]^{aa}$) to make the distributions amenable to 1D simulation.

Using $J^\alpha$, the electron fraction $Y_e$, and the baryon number density $n_b$, we can compute the neutrino and background matter contributions to $\Sigma^\alpha$ (equations \ref{eq:Sigma_neutrinos} and \ref{eq:sigma_matter}), which in turn let  us calculate components of the coherent evolution operator $H(x)$ (Equation \ref{eq:HR}) and the resonance condition (Equation \ref{eq:general_resonance}). 

\subsection{Simulating the fast-flavor instability}
\label{sec:methods_ffi}
In addition to testing for the presence of the FFI, we also directly simulate the evolution and saturation of the FFI at select locations using the particle-in-cell code Emu in one spatial dimension and two momentum direction dimensions (i.e., energy-integrated, \cite{3D_simulation}), assuming all neutrinos to be at the same arbitrary energy. Emu splits the neutrino distribution into many computational particles, each of which represents a number of physical neutrinos and antineutrinos and that encodes their flavor-space density matrices, four-position, and four-momentum. These computational particles move at the speed of light in the direction specified by their momentum, and their densities evolve according to a Hamiltonian constructed from the distribution moments accumulated onto a background grid. We resolve the instability with  512 grid cells, 1506 particles per grid cell, and periodic boundary conditions. Each cell is initialized with collection of computational particles with uniformly distributed directions of motion so the cells and particles evenly partition the phase-space of the ensemble, but the number of physical neutrinos each particle represents is varied to create an anisotropic distribution corresponding to specified number and flux densities according to the maximum entropy closure \cite{cernohorsky1994maximumentropy}. Note that the maximum entropy angular distribution is consistent with the assumption underlying the moment closure employed in the radiation transport in the original neutron star merger simulation. Since the system is rotated such that the ELN is along the dimension with spatial extent, the dominant unstable modes are able to grow, allowing these 1D simulations to accurately represent the full 3D dynamics \cite{3D_simulation,froustey2023neutrino}.

It should be noted that this approach is approximate in a few ways. First, the Hamiltonian used for the time-evolution of the simulation is different from that of Equation \ref{eq:HR}, since {\tt Emu} excludes collision and spin-coherent terms. The purpose of this approach is to probe whether there is potential interplay between the FFI and helicity coherence without directly simulating these additional terms. Second, although the FFI can induce significant flavor transformation, the ubiquity of the FFI (which tends to erase crossings) would likely transform neutrino flavor in a way that prevents crossings from being present in the first place. By analyzing local simulations of the FFI we probe the potential magnitude of the interplay between flavor transformation and helicity transformation, but self-consistent calculations are of course needed for a more robust answer.

We also note that we can also use the maximum entropy crossing condition of \cite{johns_2021_crossings,richers_2022_evaluating} to determine which regions exhibit a crossing in the electron lepton number distribution and therefore are unstable to the FFI. Specifically, the maximum entropy condition for the FFI (considering only electron neutrinos and antineutrinos) is
\begin{equation}
    \frac{\eta^2}{\alpha^2 + \gamma^2} \leq 1\,\,,
\end{equation}
where $\eta=\ln(nZ\sinh \bar{Z}/\bar{n}\bar{Z}\sinh Z)$, $\gamma = \bar{Z}\cos\theta_{\bar{F}} - Z\cos\theta_F$, $\alpha=\bar{Z}\sin\theta_{\bar{F}} - Z\sin\theta_F$, and $\theta_F-\theta_{\bar{F}}$ is the angle between the directions of the antineutrino (barred) and neutrino (unbarred) flux directions. $Z$ and $\bar{Z}$ are the parameters that describe the anisotropy of each distribution according to the maximum entropy distribution
\begin{equation}
    f(\Omega) \sim e^{Z \cos(\theta)}\,\,,
\end{equation}
where $\cos \theta=\mathbf{\Omega \cdot \hat{F}}$ is the angle between the propagation direction $\mathbf{\Omega}$ and the flux direction $\hat{F}$. Specifically, $Z$ is related to the flux and density moments as
\begin{equation}
    \frac{|\mathbf{F}|}{n} = \coth(Z)-\frac{1}{Z}\,\,. 
\end{equation}

\section{Results}
\label{sec:results}

 In Section~\ref{sec:3dspinflip}, we detail a full momentum-space analysis of spin-flip resonance and adiabaticity for a specific example point of the neutron star merger simulation of \cite{merger_simulation}. In Section~\ref{sec:FFIspinflip}, we use the data of this point as an initial condition for the fast flavor instability simulation described in \ref{sec:methods_ffi}, and see how the resultant flavor-mixing affects the resonance and adiabaticity of spin oscillations. Lastly, in Section~\ref{sec:merger} we generalize our discussion to the rest of the merger snapshot. We find the regions in which resonant helicity oscillations can be expected in some direction, and determine that these regions are usually unstable to a fast flavor instability. Finally, we describe the extent to which our analysis of resonance and adiabaticity in Sections~\ref{sec:3dspinflip} and~\ref{sec:FFIspinflip} generalizes to other locations.

\subsection{3-Dimensional Treatment of Spin Oscillations}
\label{sec:3dspinflip}
In this subsection we examine the resonance and adiabaticity of spin oscillations for neutrinos at a single example cell in the neutron star merger simulation of \cite{merger_simulation}, located at $(17.7, -21.1, -23.8)\ \mathrm{km}$. At this point the neutrino number densities are $(1.35,1.98, 0.45)\times10^{33}\,\mathrm{cm}^{-3}$ for electron neutrinos, electon antineutrinos, and each flavor of heavy lepton neutrino, respectively. The distributions of the three neutrino flavors have flux factors of $(0.26, 0.51, 0.62)$, and the flux directions yield conditions that both exhibit helicity resonance and fast flavor instability.

\subsubsection{Resonance}

We begin with an analysis of the directional dependence of resonance at the chosen cell. Recall that a necessary condition for resonance is that $\Omega\geq 1$, where $\Omega$ is the resonance parameter defined in Equation \ref{eq:Omega}. $\Omega$ depends only on the structure of the Hamiltonian (Equation \ref{eq:HR}), and specifically on the difference in magnitude of the left- and right- handed parts of the Hamiltonian's eigenvectors. Since in an anisotropic background the Hamiltonian is dependent on the direction of the neutrino momentum, the generalized resonance condition is also a function of direction.

 \begin{figure}
    \centering
    \includegraphics[width=\linewidth]{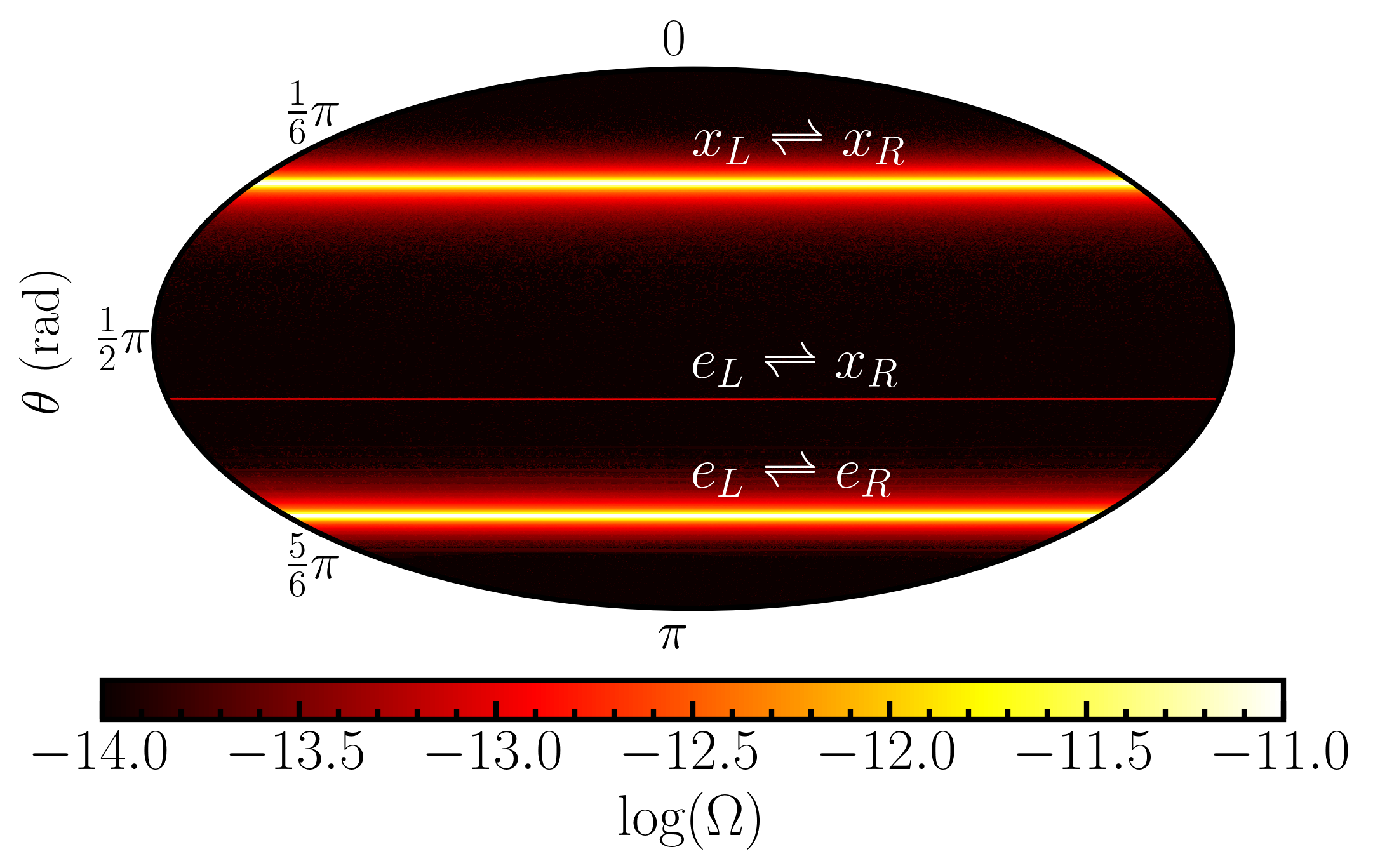}
    \caption{A Mollweide directional plot of the resonance parameter $\Omega$ (Equation~\ref{eq:Omega}) at the example location marked with a green $\times$ in Figure~\ref{fig:conditions_init}. We see three bands where oscillations are resonant, each between a particular pair of flavors as labelled in the image (with $x$ representing a general heavy lepton flavor). Note the $e_L\rightleftharpoons x_R$ is much fainter than the other two. As explained in the text, these 3 bands subdivide into a total of 9, grouped so closely that they aren't distinguishable. The polar axis is aligned with the spatial part of the left-handed electron neutrino ELN flux vector ($[J_{L}^{i}]^{ee}$.)}
    \label{fig:resonant_directions_init}
\end{figure}

Figure \ref{fig:resonant_directions_init} shows a directional logarithmic heatmap of the resonance parameter $\Omega$ for the neutrino distribution at our example point. The polar axis is parallel to the electron lepton number current $[J_L^{i}]^{ee}$. This plot presents a number of remarkable features. Firstly, we see that the resonance parameter is largely dependent only on the polar angle -- that is, on the angle between the neutrino momentum $\vec p$ and $\mathrm{Tr}_\mathrm{f}(J^{i})$. This is because the directional dependence of the Hamiltonian is dominated by the $\Sigma_{L,R}^\kappa$ terms in Equations \ref{eq:HR}, which is itself linear in the dot product between $J^{\alpha}$ and the neutrino momentum direction (see Equation~\ref{eq:Sigma_neutrinos}). Initially the only nonnegligible flavor component of $J^\alpha$ is the $ee$ part, hence $[J^{i}]^{ee}$ defines the directional dependence of $H$.

Far less intuitive features of Figure~\ref{fig:resonant_directions_init} are the azimuthally symmetric `bands' where $\Omega$ becomes large. We can identify 3 of these resonance bands in the plots, the central one far narrower than the other two. These are the only regions for which $\Omega\geq 1$, so they represent all directions exhibiting spin-flip resonance. Recall that Equation \ref{eq:general_resonance} is a necessary condition, so that not all points within the bands are guaranteed to be resonant; however, a direct inspection of directions within these bands shows that in general they all present large-amplitude, if not full amplitude oscillations. 

Interestingly, each of the bands exhibits resonant spin oscillations between a particular left-handed flavor superposition and a particular right-handed flavor superposition. For example, in Figure \ref{fig:resonant_directions_init} the bottom band labelled $e_L\rightleftharpoons e_R$ contains all directions exhibiting resonant spin oscillations between electron neutrinos, and exclusively exhibits resonance in this channel. Each of the other bands similarly shows resonance between a unique pair of flavors, labelled in Figure \ref{fig:resonant_directions_init}. Note that since the fluxes of left- and right-handed heavy lepton neutrinos are equal, the $\mu$ and $\tau$ flavors are effectively degenerate, and we can focus on a two-flavor $(e,x)$ subspace. In Section \ref{sec:FFIspinflip} we will see the flavor-mixed case where this simplification can't be made.

\begin{figure}
    \centering
    \includegraphics[width=\linewidth]{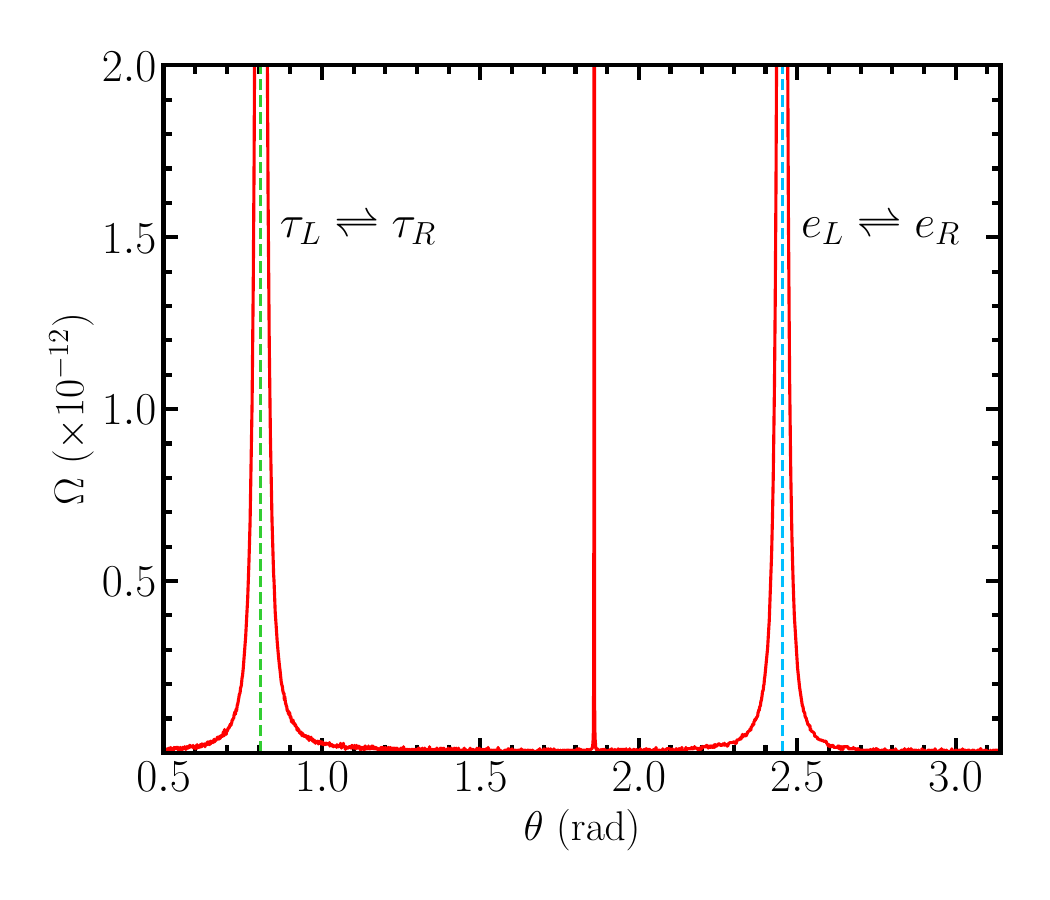}
    \caption{The resonance parameter $\Omega$ plotted against the polar angle $\theta$ at $\phi = \pi$ for the same point as in Figure \ref{fig:resonant_directions_init}. Directions satisfying the simplified resonance condition in the  $e_L\rightleftharpoons e_R$ (cyan) and $\tau_L\rightleftharpoons \tau_R$ (green) channels are indicated, and as expected they are contained in the resonance bands where $\Omega \geq 1$. The narrow central band is unique, as explained in the text.}
    \label{fig:linear_resonance_init}
\end{figure}

Figure \ref{fig:linear_resonance_init} shows $\Omega$ plotted against the polar angle $\theta$ at $\phi=\pi$, essentially displaying a vertical cross-section of Figure~\ref{fig:resonant_directions_init} so that the bands can be seen more clearly. Here we have plotted vertical dashed lines corresponding to locations satisfying the simplified resonance condition (Equation \ref{eq:resonance_simplified}) in the $e_L\rightleftharpoons e_R$ and $\tau_L\rightleftharpoons \tau_R$ channels. Recall this means these are the only locations for which resonant oscillations can occur between these flavors. These lines overlap perfectly with locations where $\Omega>1,$ exemplifying that each resonance band corresponds to spin transitions between particular flavors.

The central $x\rightleftharpoons e$ band is unique: zooming in on it reveals this band actually resolves into four `subbands', separated by less than $10^{-5}$ radians.
These are resonant in the $e_L \rightleftharpoons x_{1,R}$, $e_L \rightleftharpoons x_{2,R}$, $x_{1,L} \rightleftharpoons e_R$, and $x_{2,L} \rightleftharpoons e_R$ channels respectively, where $x_1$ and $x_2$ are specific superpositions of the heavy-lepton flavors that diagonalize the $H_L$ and $H_R$ blocks. These are the only channels that are resonant on these bands. 

The minuscule but nonzero separation of these $e\rightleftharpoons x$ 'subbands' is caused by a breakdown in our two-flavor assumption: small differences in the heavy-lepton neutrino masses lead to fine-splitting in the energies of the $e_L \rightleftharpoons x_{1,R}$ and $e_L \rightleftharpoons x_{2,R}$ conversion channels, for example, so that these are no longer entirely degenerate.

\begin{figure}
    \centering
    \includegraphics[width=\linewidth]{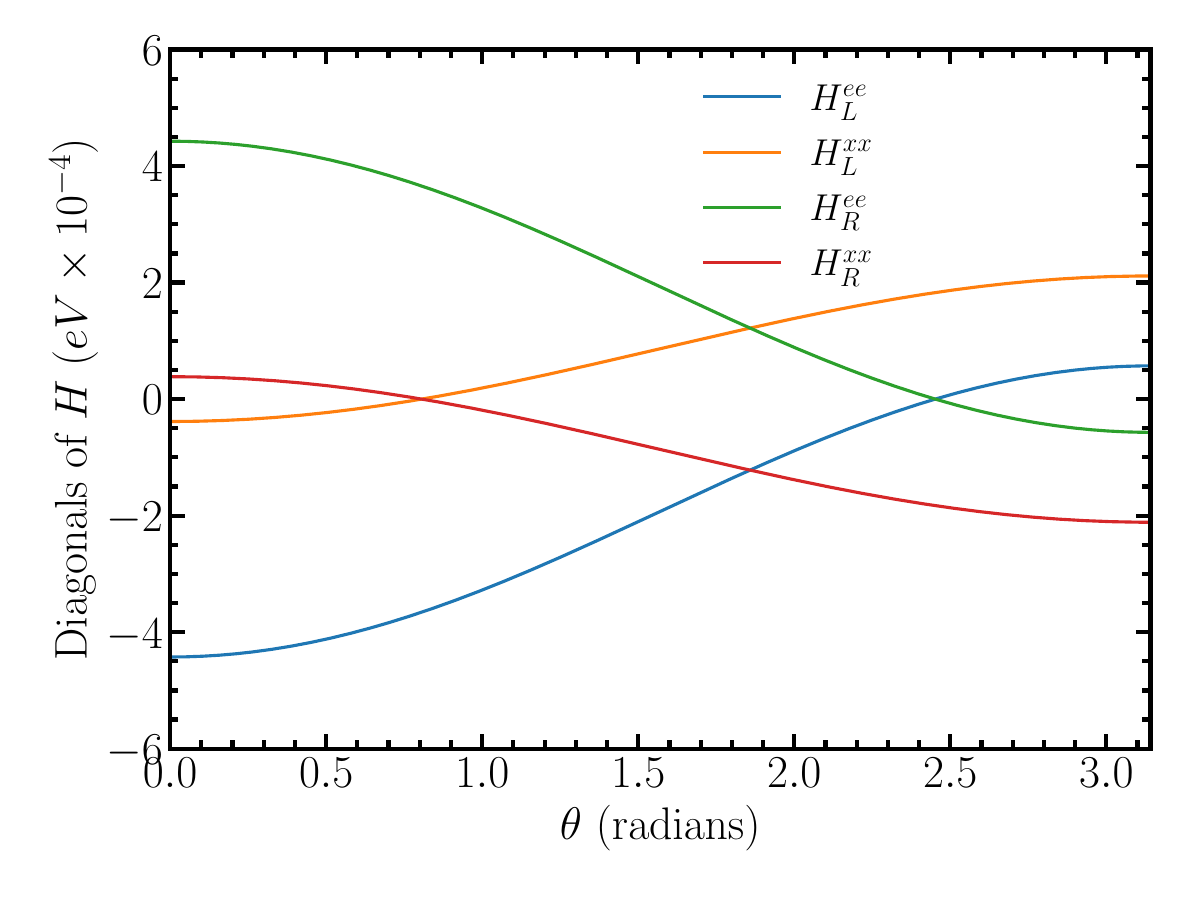}
    \caption{    
     Plot of the diagonal components of the Hamiltonian against the polar angle $\theta$ at azimuthal angle $\phi = \pi$, in the 2-flavor approximation. Due to the angular dependence of $\Sigma_L^\kappa$ and $\Sigma_R^\kappa$, components of $H_L$ and $H_R$ have a cosine-like variation with $\theta$ and tend to switch their signs as $\theta$ increases from 0 to $\pi$. Consequently the diagonals of $H_L$ and $H_R$ become equal at four locations, corresponding to the four pairs of opposite-helicity diagonals. These crossings line up perfectly with the resonance bands we see in Figures \ref{fig:resonant_directions_init} and \ref{fig:linear_resonance_init}. Note the middle two crossings between $H_{L/R}^{ee}$ and $H_{R/L}^{xx}$ overlap almost exactly, so they appear as the single faint band in Figures \ref{fig:resonant_directions_init} and \ref{fig:linear_resonance_init}.
     }
    \label{fig:H_components_polar}
\end{figure}

One might expect a similar splitting to occur for the $x_L\rightleftharpoons x_R$ resonance band, which should separate into four bands corresponding to $x_{1,L}\rightleftharpoons x_{1,R}$, $x_{1,L}\rightleftharpoons x_{2,R}$, $x_{2,L}\rightleftharpoons x_{1,R}$, and $x_{2,L}\rightleftharpoons x_{2,R}$. However, in this case the width of each of these bands is greater than their separation, so that they overlap entirely and combine into a single band encompassing helicity transitions between all heavy-lepton flavor superpositions, including $\tau\rightleftharpoons\tau$ and $\mu\rightleftharpoons\mu$.

We can use the magnitude of the flavor-diagonal components of the Hamiltonian to intuitively understand why resonance occurs at specific directions in this band structure. As suggested by the resonance condition in Equation~\ref{eq:resonance_simplified}, the resonant bands appear at angles where the $\Sigma^\kappa_{L,R}$ terms in $H_L$ and $H_R$ (Equation~\ref{eq:HR}) cancel out vacuum and matter contributions in such a way that $H_{LR}$ becomes the dominant part of the Hamiltonian. If the spatial ELN current $J^i$ is sufficiently large, then for directions near the poles in Figure \ref{fig:resonant_directions_init},  $\Sigma^\kappa_L$ and $\Sigma^\kappa_R$ will be the most important terms in Equation~\ref{eq:HR} since these terms are linear in $\vec p \cdot J^i$. (Recall the polar direction in Figure \ref{fig:resonant_directions_init} is defined to be $[J_L^{i}]^{ee}$, the maximal flavor component of $J^i$, so that $|\vec p \cdot J^i|$ is maximal for $\theta\approx 0$ or $ \pi$.) Hence for directions near the poles we can approximate $H_L\approx \Sigma^\kappa_L$ and $H_R\approx \Sigma^\kappa_R$. Also, since $\Sigma^\kappa_R = -{\Sigma^\kappa_L}^T$, we have that $H_L$ and $H_R$ will in general have opposite signs in these directions. 

From here we see that the resonance bands must arise due to the intermediate value theorem: as we move from one pole to the other, $\Sigma^\kappa_{L}$ and $\Sigma^\kappa_{R}$ will switch their signs due to the $\vec p \cdot J^i$ dependence, so the same will happen to every component of $H_L$ and $H_R$. Since $H_L \approx -H_R$ near the two poles, at some angle the components of the two matrices will cross each other and be equal. If $H_L$ and $H_R$ are approximately diagonal (as they are when flavor mixing is minimal),
then resonance bands appear at every location where a diagonal component of $H_L$ becomes equal to one of $H_R$, as in the simplified resonance condition in Equation~\ref{eq:resonance_simplified}.

Figure \ref{fig:H_components_polar} illustrates this angular dependence of $H_L$ and $H_R$, showing how the bands arise at points where the components of these matrices cross each other. All elements follow a cosine-like dependence and switch signs as $\theta$ varies from $0$ to $\pi$. Comparing to figure \ref{fig:linear_resonance_init}, it is clear that resonance bands appear where the diagonals of $H_L$ cross those of $H_R$. For example, there is an angle where $H_L^{ee} = H_R^{ee}$, which as we know implies that helicity oscillations will be resonant in the $e_L\rightleftharpoons e_R$ channel. The angle at which this happens is exactly the position of the $e_L\rightleftharpoons e_R$ resonance band, and similarly the other bands correspond to crossings between the other pairs opposite-helicity diagonal elements. Note again that we have two distinct crossings $H_L^{ee} = H_R^{xx}$ and $H_L^{xx} = H_R^{ee}$ that are separated by less than $10^{-5}$ radians, so these appear as a single band.
 
There is a constant offset to the cosine-like variation of the diagonals, which moves the resonance band crossings away from the equator. This comes from the $J^0$ contribution to $\Sigma_{L,R}^\kappa$, which adds a direction-independent shift to the potential. For the example point analyzed in this section , $J^i$ is large enough relative to $J^0$ that the diagonals of $H$ have opposite signs at the poles, so that resonance bands appear by the intermediate value theorem, but if the shift from $J^0$ were larger than the cosine-like variation from $J^i$ there would be no crossing and hence no resonance bands. This is the situation for many points in the merger which do not exhibit resonance in any direction (see Section \ref{sec:merger}).

Although resonance is present in our example cell, its significance is extremely limited by the narrowness of the bands. The directions for which $\Omega\geq 1$ collectively span a solid angle of $6.2\times10^{-7} \ \mathrm{sr}$. Interestingly, the width of the bands is determined purely by the factor of $m^\dagger/|\vec{p}|$ in Equation \ref{eq:HR}, and is the same for all points in the merger (for neutrinos of a given energy). To see why, note that the width of the bands for the transition $a_L \rightleftharpoons b_R$ is dependent effectively on two parameters. On one hand, if the simplified resonance condition applies, we can expect resonance when $|H_{LR}^{ab}|\gg |H_L^{aa} - H_R^{bb}| $ so that the larger $|H_{LR}^{ab}|$ for some pair of flavor states the broader the resonance band. On the other hand the width of a resonance band is inversely proportional to the rate of change with $\theta$ of the difference in diagonal components $H_L^{aa} - H_R^{bb}$ at the crossing where the band appears. If the derivative of this difference has a small magnitude at the direction where the components become equal, then there will be a greater angular width over which the difference is close to 0. The width of the bands can thus be approximated as
\begin{equation}
    \Delta \theta_\mathrm{resonant} \approx \tfrac{\left|H_{LR}^{ab}\right|}{\tfrac{\mathrm{d}}{\mathrm{d}\theta} \left(\left|H_L^{aa} - H_R^{bb}\right|\right) }.
    \label{eq:resonance_width}
\end{equation}
However, the factors of $\Sigma_{LR}^{\pm}$ in $|H_{LR}^{ab}|$ cancel with the factors of $\mathrm d( \Sigma^\kappa)/d\theta$ in $\mathrm d|H_L^{aa} - H_R^{bb}|/d\theta$, so that regardless of the specific distribution of the neutrino flux at each point the width of the band corresponding to any given flavor conversion channel is the same, and is of order $[m^\dagger]^{ab}/|\vec{p}|$. Unfortunately, this means we can't expect the resonant angle to get much broader in virtually any context (for neutrinos of a given energy and mass).

Incidentally this is why the central $e \rightleftharpoons x$ bands are extremely narrow compared to the $e \rightleftharpoons e$ and $x \rightleftharpoons x$ bands: $\left|H_{LR}^{ex}\right|$ is far smaller than $\left|H_{LR}^{ee}\right|$ and $\left|H_{LR}^{xx}\right|$ thanks to the linear mass dependence of $H_{LR}$.

\subsubsection{Adiabaticity}

As emphasized in previous work, for neutrino spin oscillations to be important they must be not only be resonant but also adiabatic in the sense that the oscillations must evolve on a timescale that is shorter than the timescale over which the Hamiltonian changes. Otherwise, resonance can be lost before enough time has elapsed for any helicity conversion to occur. This subsection is devoted to a study of the adiabaticity of helicity oscillations at our example cell.

Given that the Hamiltonian changes dynamically with the neutrino distribution, a full nonlinear simulation of flavor and helicity transformation is complicated and beyond the scope of this work. However, we can get a handle on the adiabaticity of oscillations via the procedures outlined in subsection \ref{sec:methods_adiabaticity}; namely we can compute the approximate timescale over which resonance will be preserved by taking gradients of the merger snapshot data, and we can compute a timescale for spin oscillations via the corresponding off-diagonal component of the Hamiltonian. We can then determine the adiabaticity of oscillations by comparing these timescales (Equation \ref{eq:adiabaticity_general}).

The timescale for an oscillation varies widely across different resonance bands. The $e_L\rightleftharpoons e_R$ and $x_L \rightleftharpoons x_R$ bands exhibit oscillations with periods of $\SI{1.4} {\milli\second}$ and $\SI{0.7} {\milli\second}$, respectively, whereas the 4 narrow $e_L\rightleftharpoons x_R$ bands all have oscillation periods of over $\SI{31} {\milli\second}$. These timescales are dependent on the size of the off-diagonal component $|\bra{\nu_L}H\ket{\nu_R}|$ correspondent to each pair of opposite-helicity states. Indeed it is no coincidence that the resonance bands with the least energetic oscillations are also the narrowest; both of these are damped by a small off-diagonal component.

This is the main bottleneck in the significance of spin-flip effects in mergers and supernovae. The off-diagonal block $H_{LR}$ is inversely proportional to the magnitude of the neutrino momentum, and it is linear in the mass matrix (Equation \ref{eq:HR}), so that it is generally extremely small. Oscillation timescales are therefore usually far too long to produce substantial helicity conversions. Several studies have searched for conditions under which electron neutrino spin oscillations are sufficiently adiabatic with $\gamma_{ee}\geq 1$ (Equation \ref{eq:adiabaticity_simplified}), to no avail \cite{Chatelain:2016xva,Tian2017}.

\begin{figure}
    \centering
    \includegraphics[width=\linewidth]{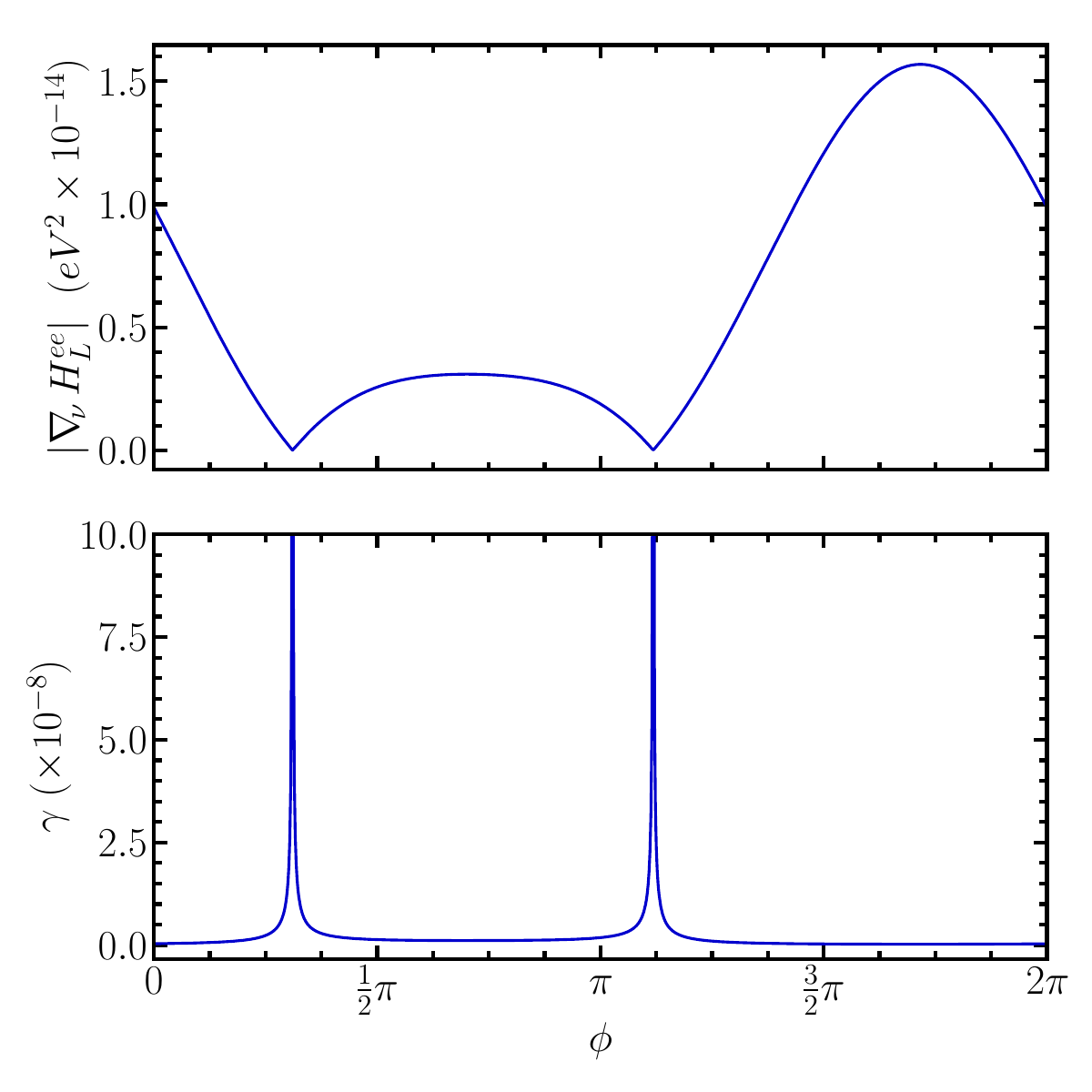}
    \caption{Gradient of the left-handed electron diagonal element of the Hamiltonian (top) and the adiabaticity $\gamma$ as defined in \ref{eq:adiabaticity_simplified} (bottom) for directions on the $e\rightleftharpoons e$ resonance band shown in Figures \ref{fig:resonant_directions_init} and \ref{fig:linear_resonance_init}, parameterized by the azimuthal angle $\phi$. We see that the gradient crosses 0 at two points, where $\gamma$ consequently grows to infinity. However, the width of directions for which $\gamma>1$ is very small.}
    \label{fig:azimuthal_init}
\end{figure}

However, we have found that incorporating the directional dependence of $\gamma_{ee}$ actually gives rise to directions that satisfy \textit{both the resonance and adiabaticity conditions}, largely due to the inverse dependence on the gradient $\nabla_\nu H_L^{aa}$. If this gradient takes different signs at two opposite directions, there will be a crossing for which it becomes arbitrarily close to 0, so that $\gamma_{ee}\geq 1$.

Figure \ref{fig:azimuthal_init} shows how $|\nabla_\nu H_L^{ee}|$ and $\gamma_{ee}$ vary along the $e_L \rightleftharpoons e_R$ resonance band highlighted in Figure \ref{fig:resonant_directions_init}, with the position on the band parameterized by the azimuthal angle $\phi$. We see that there are two points where $\nabla_\nu H_L^{ee}$ crosses 0 (kinks in the top panel), allowing the adiabaticity condition to be satisfied (peaks in the bottom panel). In these directions, significant helicity transformation is in theory possible.

The width of these crossings, however, is extremely narrow. The total azimuthal angle in Figure~\ref{fig:azimuthal_init} spanned by directions that are adiabatic is $1.0\times 10^{-9}$ rad, leading to a solid angle of around $1.0 \times 10^{-16}$ sr that is both resonant and adiabatic. Such a small angle is unlikely to contribute significantly to helicity conversion in the evolution of the merger. Nonetheless, broader adiabatic regions could appear if the overall gradient of $H_L^{ee}$ at this point had a smaller magnitude. This would make the directional derivative $\nabla_\nu H_L^{ee}$ cross 0 more gradually (top panel of Figure\ref{fig:azimuthal_init}), leading to a broader adiabatic region where $\nabla_\nu H_L^{ee}$ is sufficiently close to 0. Increasing $H_{LR}^{ee}$ would also increase the size of the adiabatic region.

Crucially, we are limited to computing derivatives using finite-width grid cell data, so this analysis assumes gradients in the background vary slowly on scales smaller than the grid cell size. By the same token, for this analysis to be valid we must demand that the timescale for a spin oscillation in some resonant direction be comparable to or smaller than the time it takes a neutrino to cross a grid cell, since our first-order gradient analysis does not give us information about changes in the background on scales much longer than that.

Unfortunately, there are no cells in the merger that can come close to satisfying this restriction. The point shown here has an oscillation lengthscale about 600 times longer than the grid cell size for resonant $e_L\rightleftharpoons e_R$ conversions, so that even along the particular direction where the adiabaticity condition is satisfied (Figure \ref{fig:azimuthal_init}), significant helicity transformation is still extremely unlikely to occur. Other locations in the merger tend to have similar or even longer oscillation length scales; see Section~\ref{sec:merger} for further details.

These results suggest that significant spin oscillations are extremely unlikely at this example point, in line with previous results. A full-momentum-space analysis of spin oscillations shows that an extremely small range of directions at any point can actually be resonant, spanning a maximum of about $4.7\times10^{-7} \ \mathrm{sr}$ for our choices of neutrino mass and energy. Our direction-dependent analysis does produce specific directions for which oscillations satisfy the adiabaticity condition (Equation \ref{eq:adiabaticity_general}), but these range about $10^{-16} \ \mathrm{sr} $, and neutrinos travelling along these directions are still unlikely to experience significant helicity transition due to the extremely long length scales associated with this effect.

\subsection{Effect of the Fast Flavor Instability}
\label{sec:FFIspinflip}

\begin{figure}
    \centering
    \includegraphics[width=\linewidth]{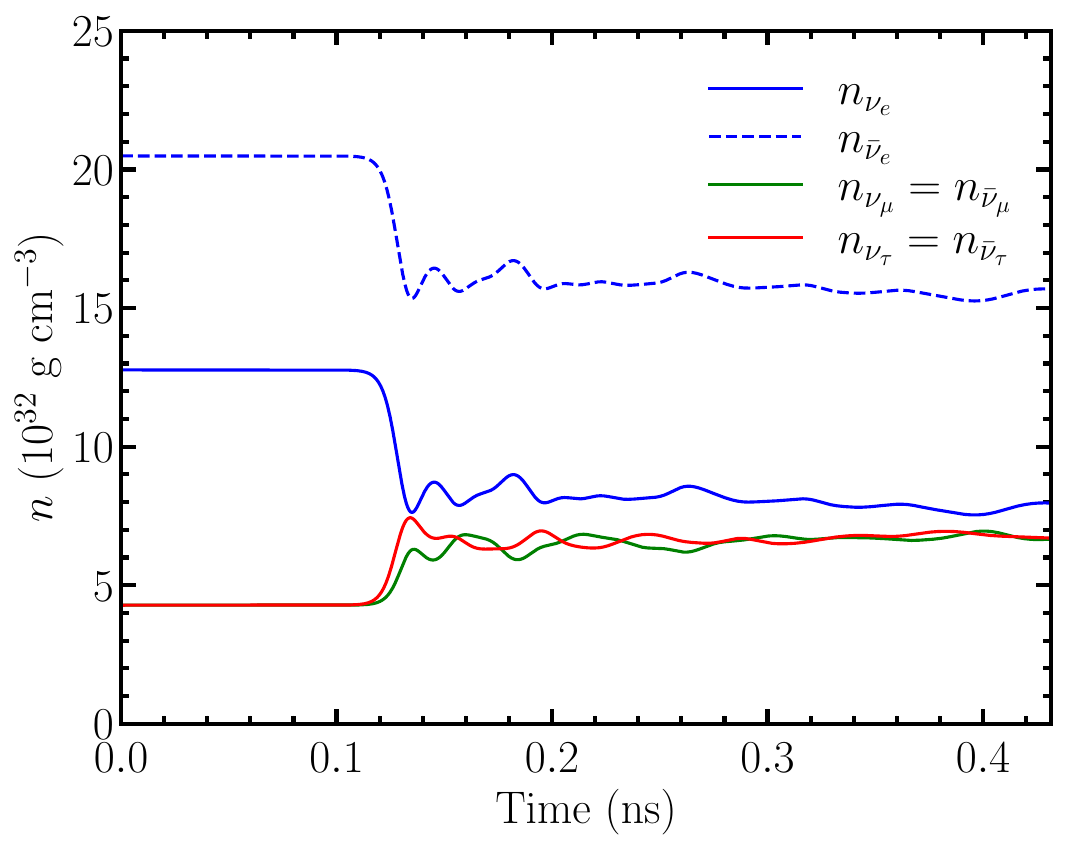}
    \caption{Evolution of the number density of each neutrino flavor due to the fast flavor instability. There are initially more electron neutrinos (solid blue) than electron antineutrinos (dashed blue) and heavy lepton neutrinos (green and red). In this case, by the end of the simulation there are similar numbers of all neutrino types, although this is not always true. Not shown are the changes in the magnitudes and directions of the flux of each neutrino flavor. }
    \label{fig:flavorcange}
\end{figure}

\begin{figure}
    \centering
    \includegraphics[width=\linewidth]{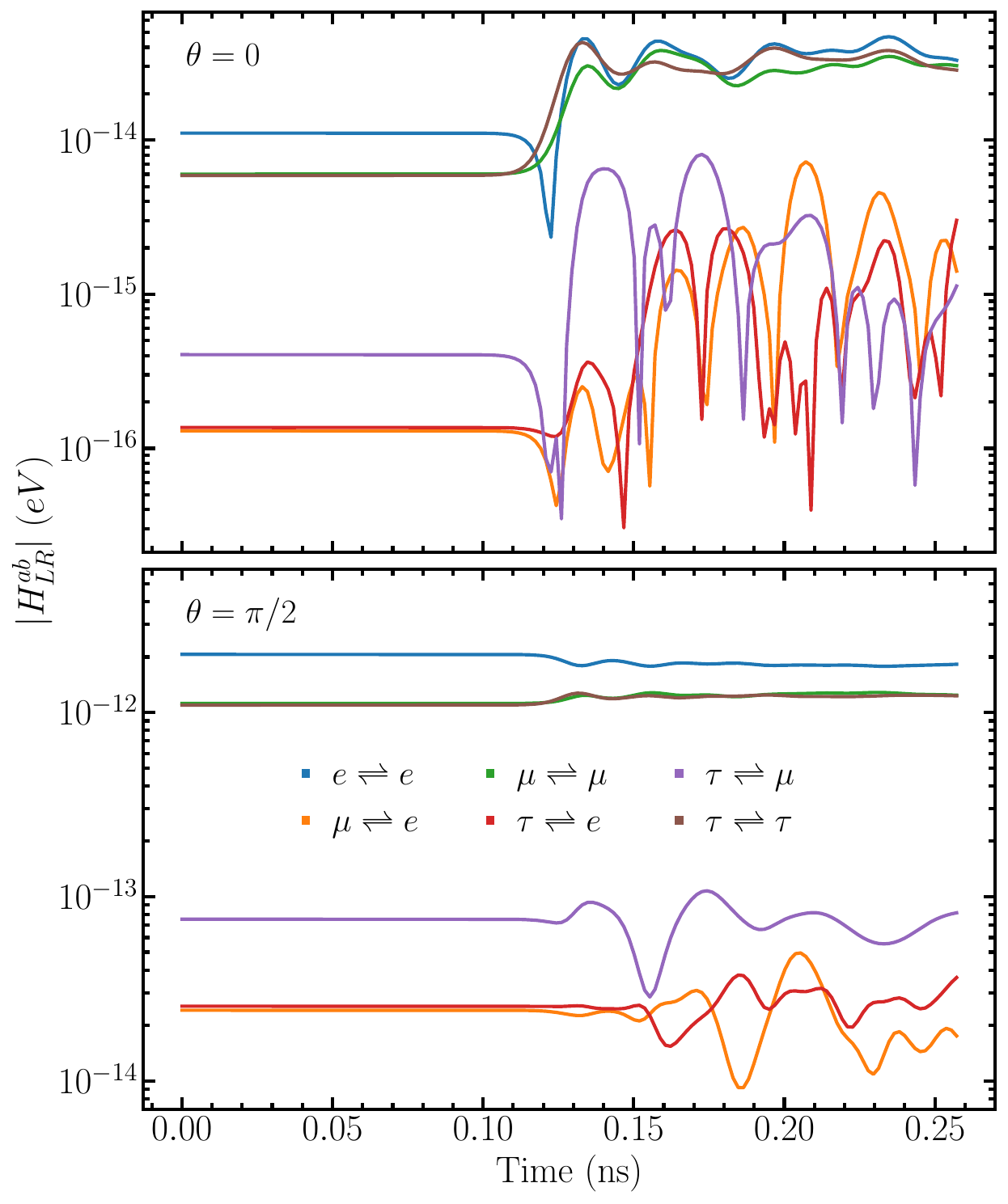}
    \caption{Change in $|H_{LR}^{ab}|$ during the FFI at our sample point, along a direction initially parallel (top) and perpendicular (bottom) to $|J^i|$ (polar direction in Figure~\ref{fig:resonant_directions_init}). The change in the helicity-flip Hamiltonian depends on the component and the neutrino direction. The legend labels components of $H_{LR}$ according to the flavor channel they correspond to.}
    \label{fig:hlr_time_evolution}
\end{figure}
In regions unstable to the FFI, the neutrino distributions quickly evolve to erase any existing ELN crossings. The net result on the total number of each flavor at our test point is displayed in Figure~\ref{fig:flavorcange}. After the FFI saturates, the total numbers of electron neutrinos and antineutrinos decreases while the number of heavy lepton neutrinos increases. These transformations proceed in a way that preserves $n_{\nu_a}-n_{\bar{\nu}_a}$ for each neutrino flavor $a$, since we only consider the neutrino self-interaction Hamiltonian here. Even though the net lepton number is conserved for each flavor, the net flux is not, and the magnitude and direction of the flux vectors change significantly. 

\begin{figure}[t]
    \centering
    \includegraphics[width=\linewidth]{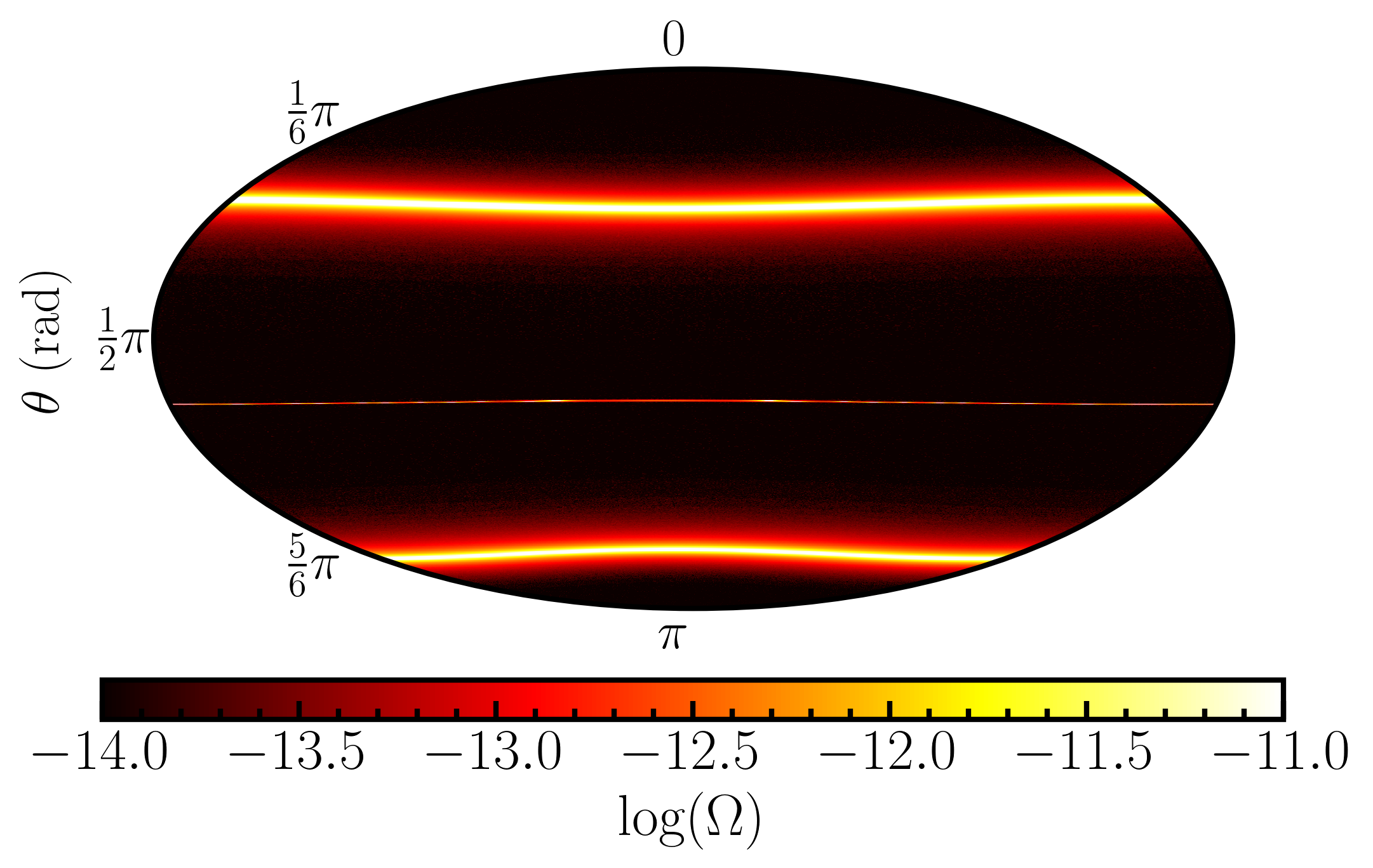}
    \caption{The same directional plot of the resonance parameter $\Omega$ (Equation \ref{eq:Omega}) as in Figure \ref{fig:resonant_directions_init}, and at the same location in the merger, but after the fast flavor instability has transpired. The polar direction has not changed -- it is the pre-instability direction of $\mathrm{Tr}_\mathrm{f}[J^i]$.
    The resonance bands seen initially in Figure \ref{fig:resonant_directions_init} have separated into several more visible bands, and these no longer lie symmetrically around the polar direction. }   
    \label{fig:resonant_directions_final}
\end{figure}

\begin{figure*}[ht!]
    \centering
    \includegraphics[width=\linewidth]{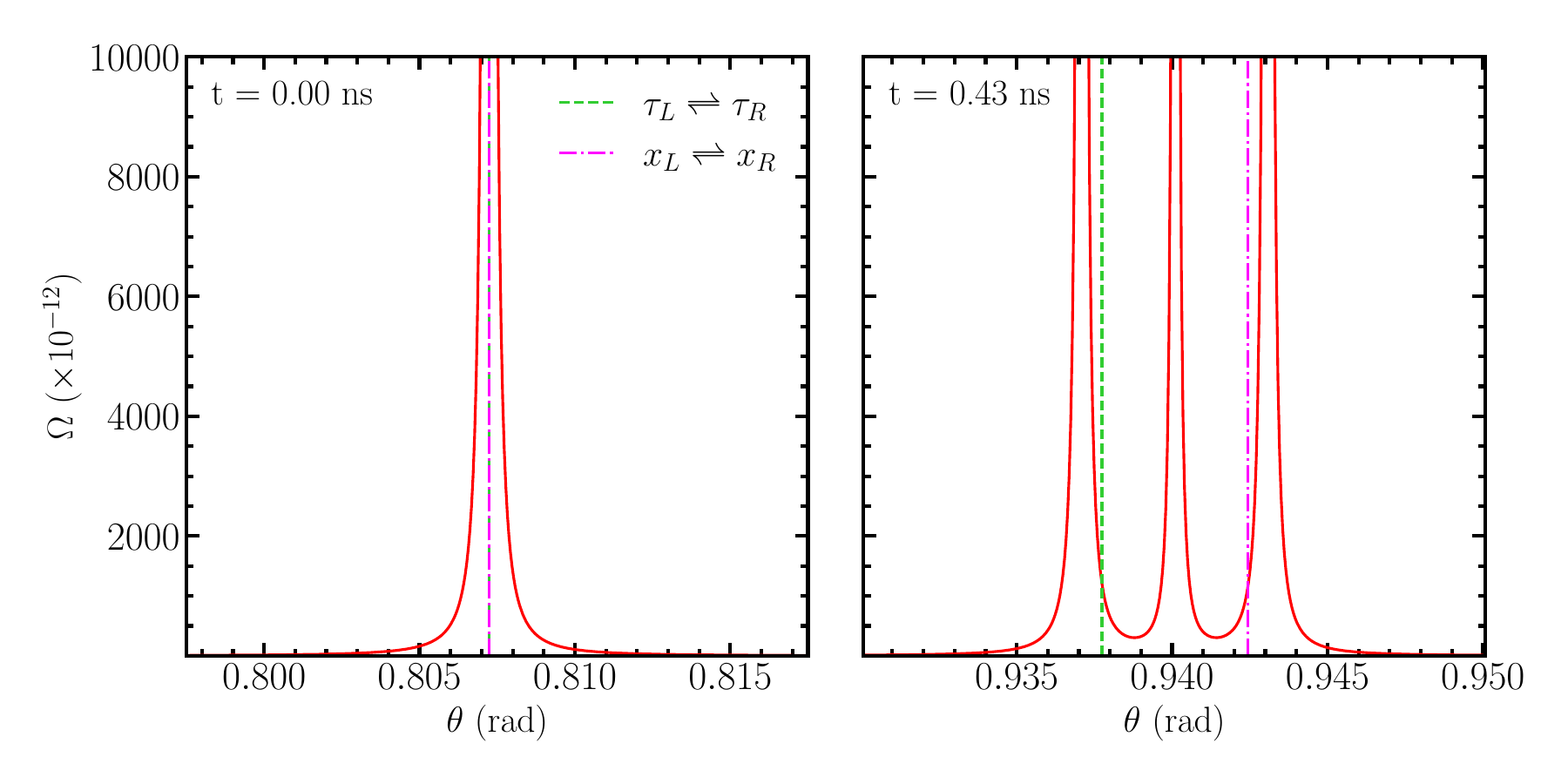}
    \caption{A zoomed-in version of Figure \ref{fig:linear_resonance_init}, showing the heavy lepton $x_{L}\rightleftharpoons x_{R}$ resonance band before (left) and after (right) the fast flavor instability. We plot $\theta$ against the resonance parameter $\Omega$ at azimuthal angle $\phi = \pi$, as in Figure \ref{fig:linear_resonance_init}. After the instability the heavy-lepton bands, which were previously overlapping, separate into four subbands (the middle two are so close together that they look like one band in the right panel, but they are in fact separate.)
    Directions satisfying the simplified resonance condition in the $\tau\rightleftharpoons \tau$ and $\mu\rightleftharpoons \mu$ channels are indicated. Pre-instability, both these directions line up perfectly with the overlapping bands so that oscillations are resonant in these channels. However, flavor mixing causes the degeneracy in heavy lepton flavors to break down so that these directions no longer match up with a resonant region as seen in the right panel; after the instability there is no resonance in the $\tau\rightleftharpoons \tau$ and $\mu\rightleftharpoons \mu$ channels and the simplified condition breaks down. The generalized conditions still finds the resonant directions.
    }
    \label{fig:linear_resonance_zoomed}
\end{figure*}

The helicity-off diagonal components of the Hamiltonian in Equation~\ref{eq:HR} are sensitive to these changes. In Figure \ref{fig:hlr_time_evolution}, we demonstrate the time evolution of all components of $H_{LR}$ along a direction parallel to and a direction perpendicular to the initial net neutrino current $\mathrm{Tr}_\mathrm{f}[J^i]$. We see that the instability leads to substantial and direction dependent changes in the magnitude of different $H_{LR}$ components. These changes are most pronounced for directions nearly parallel or antiparallel to $\mathrm{Tr}_\mathrm{f}[J^i]$. In any case these transformations are crucial to the significance of spin oscillations at a point since, as discussed in the previous section, the magnitudes of the elements $H_{LR}^{ab}$ are factors in the angular width of the directions that exhibit spin resonance. 

The directions and conversion channels for which spin oscillations are resonant are also affected by flavor-mixing in the FFI. The post-instability resonant directions for our example point are shown in Figure~\ref{fig:resonant_directions_final}, at a timestep of 0.43 ns. As in Figure~\ref{fig:resonant_directions_init}, resonance is indicated by a large value of $\Omega$ (Equation~\ref{eq:general_resonance}). We still have a band structure, but due to the change in magnitude and direction of the net flux, the bands have rotated and moved in such a way that they are no longer symmetric around the polar axis.

Some of the bands have also separated from each other. Recall that in general we should have 9 resonance bands spanning 9 possible orthogonal flavor conversion channels, which in Figure \ref{fig:resonant_directions_init} were clumped together into 3 closely-spaced or overlapping groups of bands due to degeneracy in the heavy-lepton flavors.  This degeneracy is slightly broken as a result of different heavy lepton flavors evolving independently in the simulations of the FFI, separating out  different flavor conversion channels.

For example, in the pre-instability case the heavy lepton $x_{L}\rightleftharpoons x_{R}$ band in Figure \ref{fig:resonant_directions_init} corresponded to four resonance bands that overlapped entirely, so that any helicity conversion between heavy lepton states was resonant there. After the instability, this band has split into four distinct sub-bands as shown in Figure \ref{fig:linear_resonance_zoomed}. (note two of the sub-bands are so close together they look like a single band in \ref{fig:linear_resonance_zoomed}, however they don't overlap completely.) Each of these is resonant between two heavy-lepton flavor superpositions. Hence, while prior to the instability resonant oscillations were possible between any pair of heavy lepton states, after the instability these are only possible in four distinct channels at each of the four separated bands. 

Flavor mixing also causes a breakdown in the simplified resonance condition. To illustrate this, in Figure \ref{fig:linear_resonance_zoomed} we plot the locations satisfying the simplified resonance condition in the $\tau_L\rightleftharpoons\tau_R$ and $\mu_L\rightleftharpoons\mu_R$ channels. before the instability, these lie inside the resonance band with $\Omega \geq 1$; after the instability these locations lie between the split sub-bands, hence resonant $\tau_L\rightleftharpoons\tau_R$ and $\mu_L\rightleftharpoons\mu_R$ conversions become impossible. This demonstrates the relative advantage of the generalized condition $\Omega \geq 1$, which applies even in the flavor-mixed case and encompasses all flavor conversion channels simultaneously.

The width of the resonance bands changes due to the FFI, but marginally. Post-instability the resonant solid angle decreases to  $4.7\times10^{-7}$ steradians. This change is caused by the splitting in the heavy-lepton conversion band, which was broadened in the pre-instability case because of the overlapping in the constituent subbands. In general, resonance bands have a fixed width dependent purely on $[m^\dagger]^{ab}/|\vec p|$ (Equation \ref{eq:resonance_width}).

\begin{figure}
    \centering
    \includegraphics[width=\linewidth]{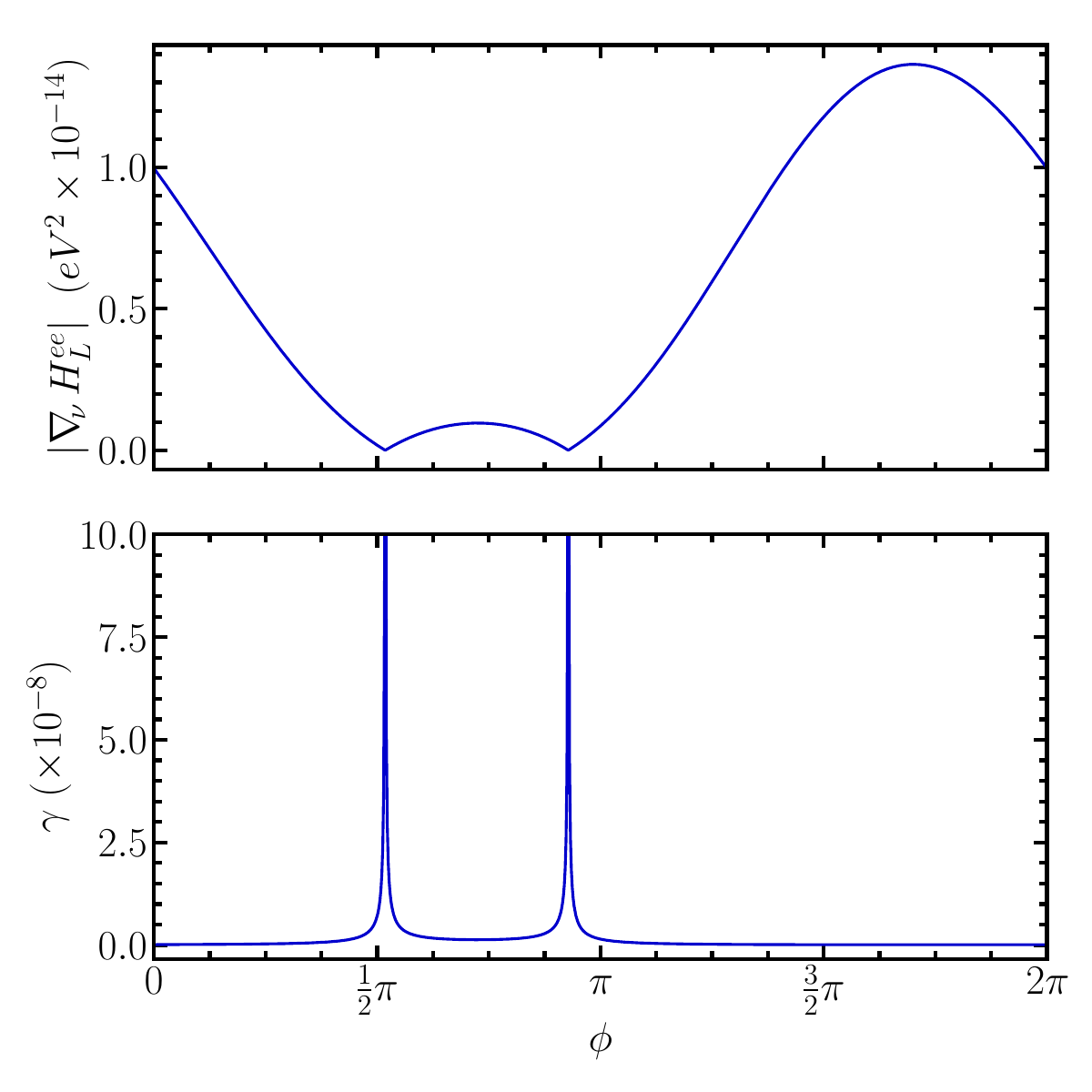}
    \caption{Gradient of $H^{ee}_L$ (top) and the adiabaticity $\gamma$ as (bottom) along the $e\rightleftharpoons e$ resonance band after a fast flavor instability. Comparing with the pre-instability case, Figure \ref{fig:azimuthal_init}, we see that the entire distribution of $H^{ee}_L$ values has shifted upwards so that the crossings are closer together and wider as they occur at a flatter part of the sinusoid. }
    \label{fig:azimuthal_gradients_final}
\end{figure}

Just as before the instability, there are crossings in the gradient that cause certain directions to become adiabatic. The variation of the adiabatic index with the azimuthal angle across the $e_L\rightleftharpoons e_R$ resonance band is shown in Figure \ref{fig:azimuthal_gradients_final} (compare to Figure \ref{fig:azimuthal_init}). We see that the sinusoid dependence of the gradient $\nabla_\nu H^{ee}_L$ has shifted upwards, so that the angles where it crosses 0 are closer together. 

The total adiabatic and resonant solid angle is still about $10^{-16}$ sr, but this upward shift of the gradient suggests a possible way of getting a broader angle: if the shift were larger, so that the tip of the sinusoid just barely touched 0, the crossing would be much broader as it would occur at a shallower slope. Of course, if the shift were larger still, there would be no crossing at all and therefore no adiabatic direction. We will see in \ref{sec:merger} that this happens at several locations.

All in all, flavor transformation effects do not seem to significantly change the total solid angle spanned by directions that are resonant and adiabatic, although the exact distribution of these directions -- and the flavor channels in which resonant oscillations occur -- is changed. This adds another barrier to helicity oscillations, since neutrinos travelling along resonant directions will stop being resonant if a fast flavor instability shifts the resonance bands, which can happen -- as it did for this cell -- in less than half a nanosecond.

\subsection{Spin Oscillations Throughout The Neutron Star Merger}
\label{sec:merger}

In this subsection we generalize the analysis of the single point described in the previous subsections to the data of the entire merger snapshot. First, we detail where the conditions for resonant spin oscillations and for a fast flavor instability are satisfied. Then we generalize the 'resonant band' analysis from previous subsections, showing how the shape and position of these bands vary at different locations. Next we show the magnitude of the resonant and adiabatic solid angle for a subsection of the merger and show how this is affected by the fast flavor instability. Finally we demonstrate that there is a only limited region where spin oscillation length scales are short enough that the adiabaticity condition of Equation \ref{eq:adiabaticity_general} can be faithfully employed. 

\subsubsection{Conditions for Flavor instability and Helicity Resonance}

\begin{figure*}[ht!]
    \centering
    \includegraphics[width=\linewidth]{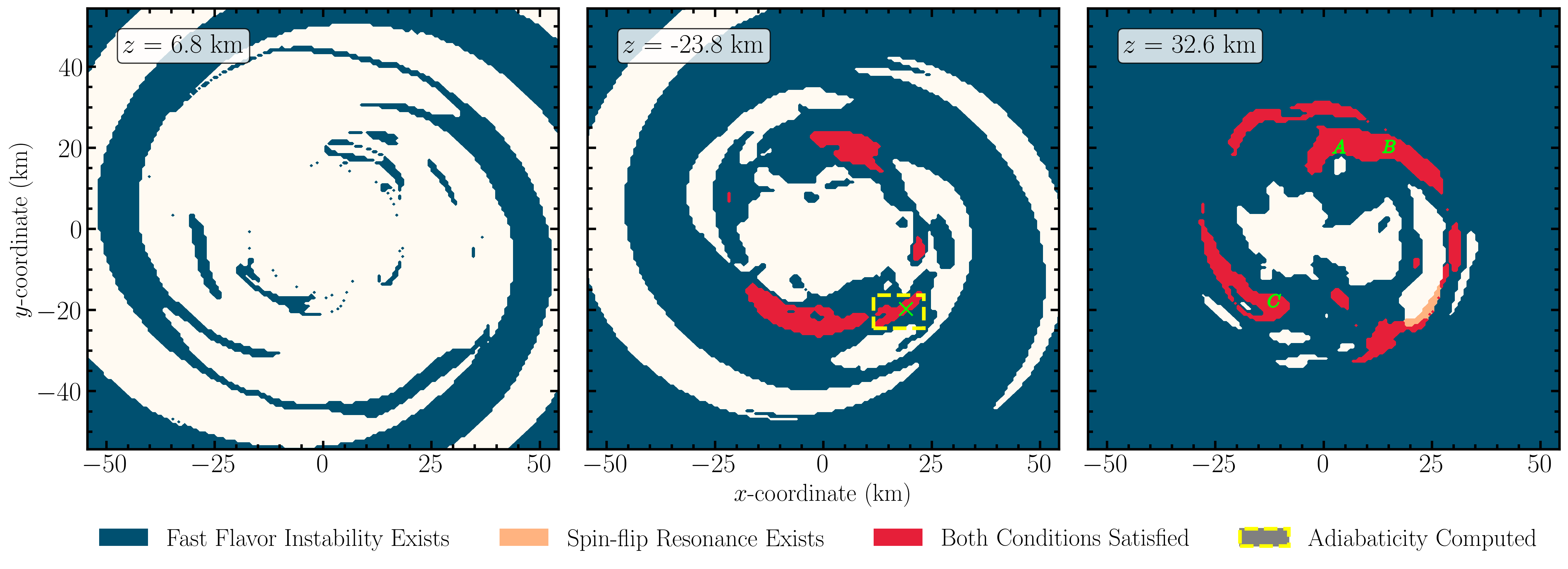}
    \caption{ Regions where the $e_L\rightleftharpoons e_R$ spin-flip resonance condition (Equation~\ref{eq:resonance_simplified_e}) and the maximum entropy angular ELN condition for a fast-flavor instability \citep{richers_2022_evaluating} are satisfied, shown for 3 cross-sections of the neutron star merger simulation at different values of the $z$-coordinate. Blue points satisfy only the fast-flavor instability condition, orange points satisfy only the spin-flip resonance condition, and red points satisfy both simultaneously, indicating the possibility for interactions between the two effects. We see most points that are resonant satisfy the fast flavor instability condition, so that orange points are relatively scarce: it is rare that spin oscillations can be analyzed in absence of flavor-mixing. Green letters represent the points for which we show resonance bands in Figure~\ref{fig:angular_resonance_multipoint}; the $\times$ marker represents the example point we discussed in Sections \ref{sec:3dspinflip} and \ref{sec:FFIspinflip}.
    The yellow square indicates the area where we will run our adiabaticity analysis later in this section. }
    \label{fig:conditions_init}
\end{figure*}

In Figure \ref{fig:conditions_init} we highlight regions that have at least one direction satisfying the simplified resonance condition in the $e_L\rightleftharpoons e_R$ channel (Equation \ref{eq:resonance_simplified_e}), as well as regions satisfying the ELN crossing condition for a fast flavor instability, for three cross-sections of the neutron star merger simulation. The left cross-section is taken near the center of the merger (which is at $z=0$); the center and right cross-sections are taken near the poles. Note we can use the simplified resonance condition since the pre-instability Hamiltonian at all locations in the merger is approximately flavor diagonal (up to mixing in the heavy lepton flavors).

As reported in \cite{grohs2023}, the fast-flavor instability is extremely pervasive (blue and red) especially near the poles where it is expected to occur almost everywhere. The prevalence of flavor instabilities exemplifies the importance of considering spin oscillations in the mixed-flavor case.

Additionally, we see broad regions for which $e_L \rightleftharpoons e_R$ spin oscillations are resonant in some direction (orange and red), and these resonant regions almost universally also satisfy the condition for a fast flavor instability (red) with only a small region in the right panel satisfying helicity resonance exclusively. Any location in these regions presents an $e_L\rightleftharpoons e_R$ resonance band similar to that of Figure \ref{fig:resonant_directions_init}. We see that resonant spin oscillations are common at the poles but are rarely present closer to the center of the merger. Note that points not in the red or orange regions could still present resonance in other flavor channels.

\subsubsection{Resonance Bands}
\begin{figure}
    \centering
    \includegraphics[width=\linewidth]{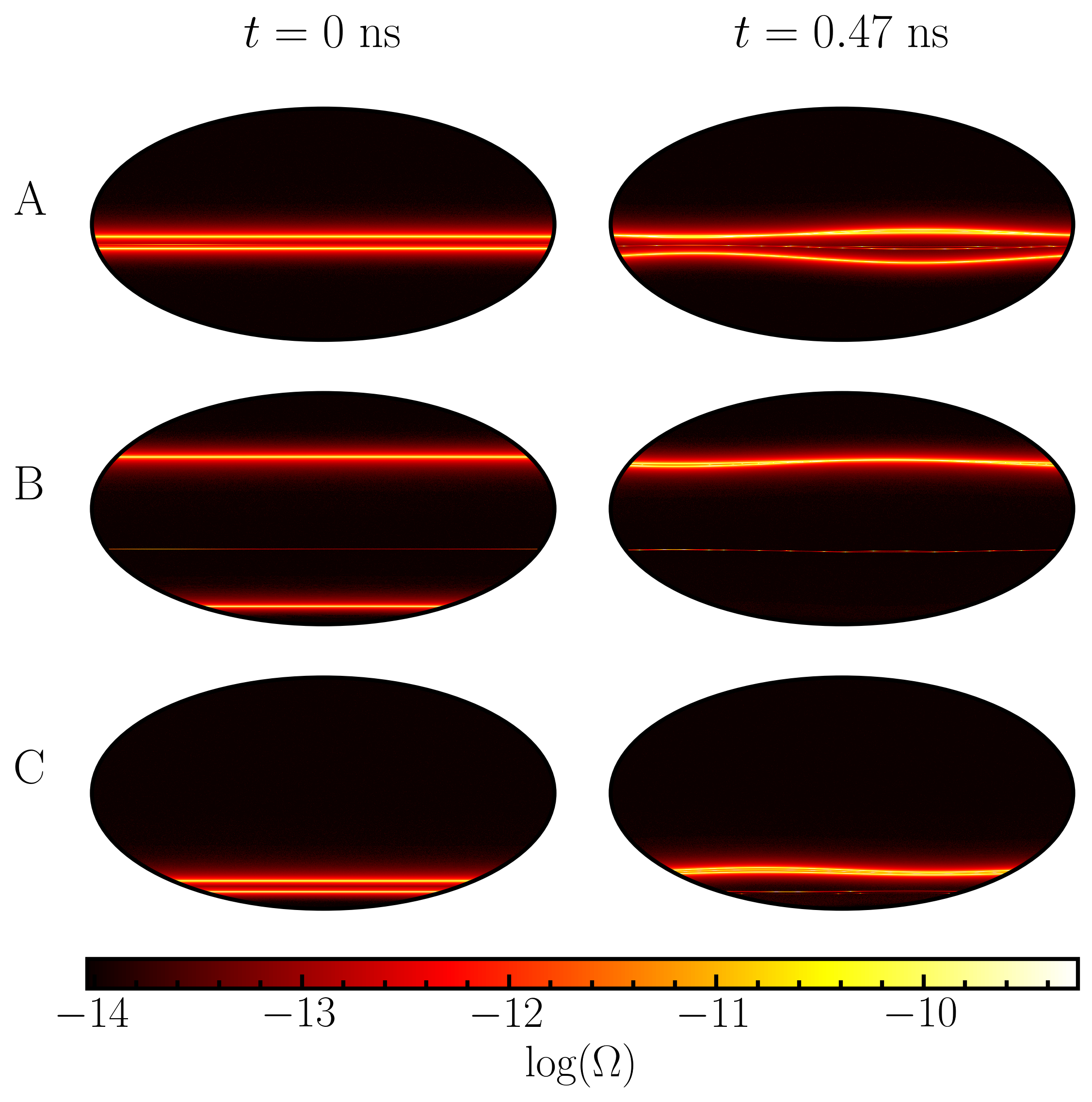}
    \caption{Resonance bands before the fast flavor instability (left) and after (right) for three other points in the merger at locations marked by letters in Figure \ref{fig:conditions_init}. We see that the initial polar angle at which the bands are located varies for different locations, and the exact splitting and rotation of the bands due to the instability varies as well. In general, points that are have resonant directions always have those directions distributed in this azimuthally symmetric structure, with a maximum of 9 (possibly overlapping) bands. }
    \label{fig:angular_resonance_multipoint}
\end{figure}

Figure \ref{fig:angular_resonance_multipoint} shows the pre- and post-instability resonant bands at 3 other locations in the merger for comparison with the example point of subsections \ref{sec:3dspinflip} and \ref{sec:FFIspinflip}. The locations of these 3 points, as well as the location of our original example cell, are marked with green letters in figure \ref{fig:conditions_init}. 

Figure \ref{fig:angular_resonance_multipoint} shows that the specific angular position of the resonance bands is highly location-dependent, as is the effect of the fast flavor instability. Points can have resonance bands at any polar angle, and their separation can be as small as $0.15$ radians (point C) or they can span the entire range of polar angles (point B). Close to the boundaries of the red resonant areas seen in Figure \ref{fig:conditions_init}, the $e_L\rightleftharpoons e_R$ band tends to get closer to the poles, and ultimately vanishes for non-resonant (blue/white) locations. Other resonance bands are almost never present far away from the red regions with $e_L\rightleftharpoons e_R$ resonance.

Interestingly, the fast flavor instability can cause resonance bands to get pushed off the spectrum entirely, as in point B where resonance in the $e_L \rightleftharpoons e_R$ channel is lost. We expect the FFI could also cause resonance to appear at a point where it was absent, though we couldn't find any points that exemplified this.

Note the width of these resonance bands is almost the same for each plot and only changes due to either one of the bands being absent or overlapping bands separating out due to the FFI as in Figure \ref{fig:linear_resonance_zoomed}. A band encompassing resonance between flavors $a$ and $b$ has a width around $[m^\dagger]^{ab}/|\vec p|$ regardless of the background matter and neutrino distribution. In any case, we never see a total resonant angle greater than $10^{-6}$ sr for neutrinos of our selected mass and momentum.

\subsubsection{Resonant and Adiabatic Directions}

We saw in section \ref{sec:3dspinflip} that for our example point there were specific directions that were resonant and also satisfied the adiabaticity condition of Equation \ref{eq:adiabaticity_general}. We now determine if other points in the merger present such directions and what range of resonant and adiabatic solid angles we can expect, before and after a fast flavor instability.

\begin{figure*}
    \centering
    \includegraphics[width=\linewidth]{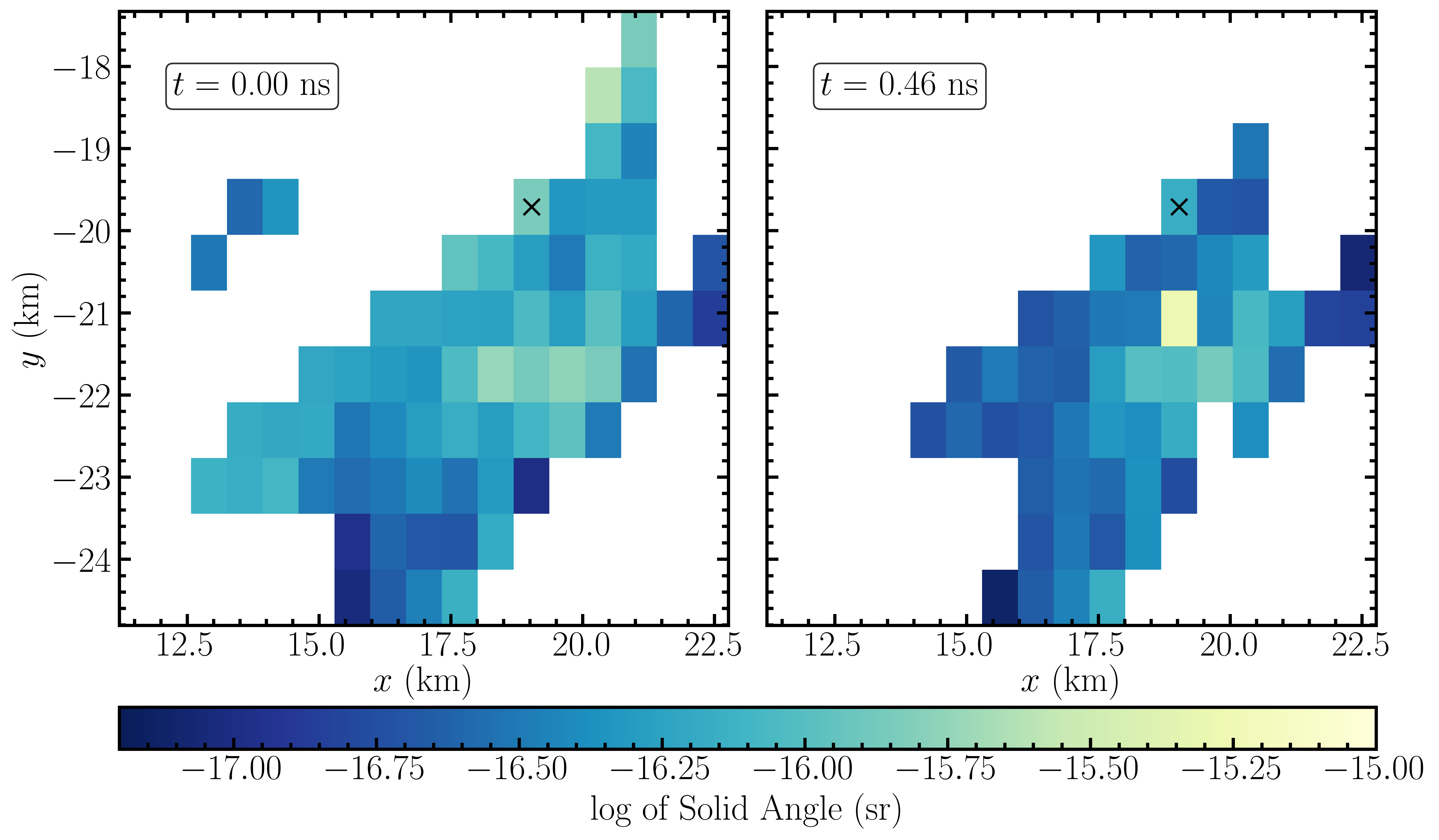}
    \caption{Total resonant and adiabatic solid angle for the section of the merger highlighted in a dashed yellow square in Figure \ref{fig:conditions_init}, before and after the fast flavor instability. The example point of sections \ref{sec:3dspinflip} and \ref{sec:FFIspinflip} is marked with a black x. We see that the fast flavor instability causes a decrease in the size of the resonant and adiabatic angle for most points in this region, and often makes this angle disappear entirely.
}
    \label{fig:adiabaticities_init}
\end{figure*}

Due to the computational intensity of finding the span of these resonant and adiabatic directions we limit our analysis to the cross-sectional region highlighted in a yellow dashed rectangle in Figure \ref{fig:conditions_init}. Figure \ref{fig:adiabaticities_init} shows the total resonant and adiabatic solid angle for cells within this region, before and after a fast flavor instability. We see that almost all cells present solid angles between $10^{-15}$ and $10^{-17}$ sr, and that the values tend to decrease after the FFI. 

Note that many points lose their resonant and adiabatic directions entirely after the instability. This can either be due to the resonant bands shifting past the poles or, more usually, due to the $\nabla_\nu H^{ee}_L$ distribution shifting so far up that it no longer crosses 0 at any point (compare Figures \ref{fig:azimuthal_init} and \ref{fig:azimuthal_gradients_final}). It is not clear if the tendency for the solid angle to decrease or disappear is unique to this small region or if it is a general quality of the FFI. Note there are points that increase their solid angle substantially.

Regardless, as we determine in the next subsection, this analysis is overturned by the fact that the timescale for an oscillation is so large that even directions satisfying the adiabaticity condition are unlikely to present significant oscillations.

\subsubsection{Spin-oscillation length scale}

The off-diagonal components of the Hamiltonian $H_{LR}$ determine the wavelength on which neutrinos change helicity, and if these wavelengths are small compared to the length scales over which the background changes then the neutrinos can be expected to traverse the resonance adiabatically, so that a large amount of helicity transformation can be expected. 

\begin{figure}
    \centering
    \includegraphics[width=\linewidth]{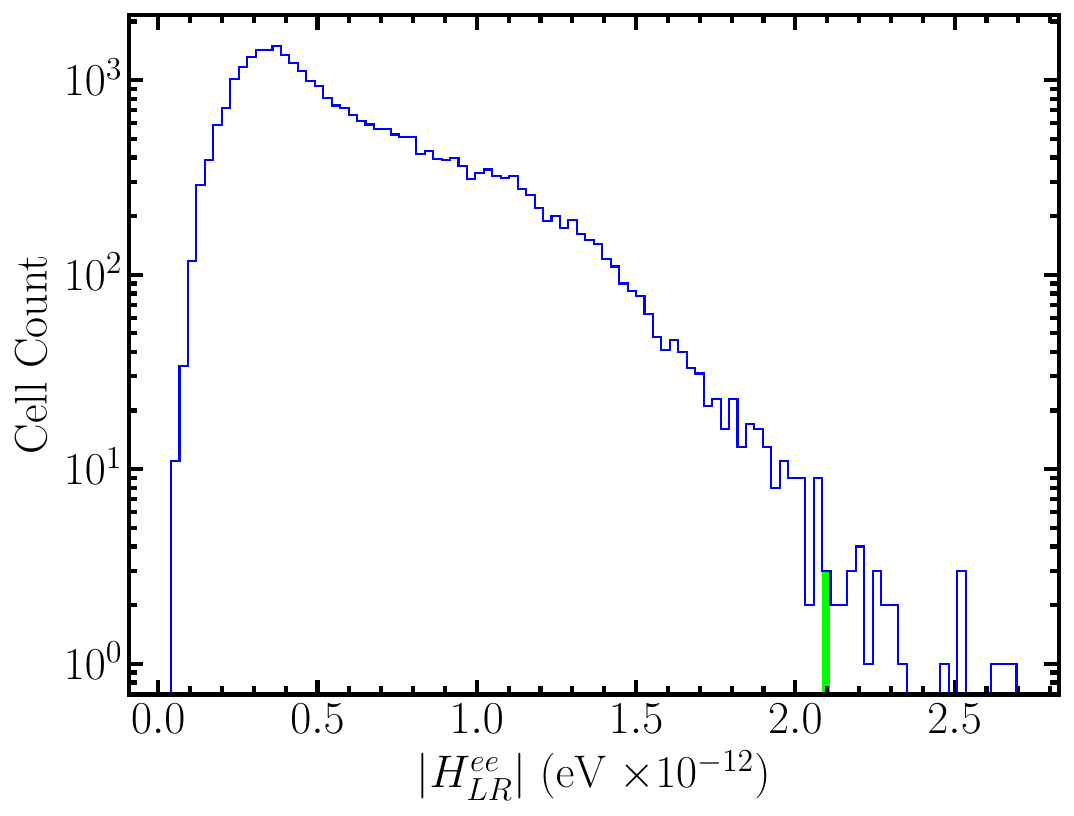}
    \caption{A logarithmic histogram plot of the magnitude of $H_{LR}^{ee}$, the helicity-off-diagonal component channelling electron neutrino spin oscillations, for the data of all cells in the merger presenting a resonance band. The green band contains the cell we chose as our example point in Sections \ref{sec:3dspinflip} and \ref{sec:FFIspinflip}. A wavelength twice the size of a grid cell corresponds to an energy of 4.6$\times 10^{-10}$ eV; cells presenting energies much lower than that cannot be used in our adiabaticity analysis.}
    \label{fig:hlr_histogram}
\end{figure}

 In Figure \ref{fig:hlr_histogram} we show a logarithmic histogram of the magnitude of $H_{LR}^{ee}$ for points in the merger presenting resonance in the $e_L \rightleftharpoons e_R$ channel (red and orange areas in Figure \ref{fig:conditions_init}). The bin highlighted in green contains the example point of sections \ref{sec:3dspinflip} and \ref{sec:FFIspinflip}, which we picked to have a particularly large $H_{LR}^{ee}$. However, even this point has an oscillation length scale of 415 km in the $e_L \rightleftharpoons e_R$ channel, an order of magnitude longer than the radius of the entire merger, and almost all other locations  will have even longer scales. 
 
 The upshot is that even for directions satisfying both the resonant and adiabatic conditions, significant oscillations are virtually impossible as the oscillations will transpire on far too long a distance. The standard adiabaticity condition of \ref{eq:adiabaticity_simplified} is not an accurate signaler of significant oscillations in this case because it only accesses local, first-order changes in the background.

\section{Conclusions}
\label{sec:conclusions}

We search for conditions conducive to Majorana neutrino helicity oscillations in a three-dimensional general-relativistic snapshot of a neutron star merger that occur purely due to anisotropy and inhomogeneity of the matter and neutrino distributions. Assuming neutrino masses within experimental bounds, we corroborate previous studies based on simpler models by demonstrating that very little helicity transformation is likely to take place. We find spin oscillation length scales to be on the order of several hundred kilometers, longer than the diameter of the entire merger, even following the effects of the neutrino fast flavor instability.

We construct a metric for helicity transformation resonance (Equation~\ref{eq:general_resonance}) that is applicable for a general mixture of neutrino and antineutrino distributions of all three flavors, and demonstrate that it simplifies to a previously proposed resonance metric when heavy lepton neutrinos and antineutrinos have identical distributions. We also estimate the adiabaticity of neutrinos passing through such a resonance for an arbitrary combination of left and right handed states (Eqaution~\ref{eq:adiabaticity_general}). This treatment allows us to estimate the prevalence of helicity transformation after the fast flavor instability mixes the neutrino distributions.

At each location in the merger we identify the neutrino propagation directions along which neutrinos experience helicity transformation resonance (e.g., Figure~\ref{fig:resonant_directions_init}) and identify a series of directional bands that each correspond to a particular helicity transformation mode. Although a significant fraction of the volume in the polar regions of the merger contains some direction along which there is resonance (Figure~\ref{fig:conditions_init}), the angular width of the resonant regions, and therefore the amount of neutrinos capable of experiencing resonance, is extremely slim at around $5\times 10^{-7} \ \mathrm{sr}$. This angular width is is the same everywhere in the merger, since it is only a function of the $m/|\vec p|$ ratio. This adds yet another barrier to the feasibility of significant helicity oscillations in mergers without additional contributions from other effects like extreme magnetic fields.

Previous work has demonstrated that even if neutrinos experience resonance, they do so non-adiabatically such that no significant helicity transformation occurs. We demonstrate that along the resonant direction bands (e.g. in Figure~\ref{fig:resonant_directions_init}) there is often some direction along which the adiabaticity condition, Equation~\ref{eq:adiabaticity_general}, is satisfied (Figure~\ref{fig:azimuthal_init}). However, the solid angle of directions that are simultaneously adiabatic and resonant is only on the order of $10^{-15}$ steradians. In addition, even those directions that satisfy Equation~\ref{eq:adiabaticity_general} have oscillation length scales larger than relevant dynamical scales in the neutron star merger (Figure~\ref{fig:hlr_histogram}), indicating that local adiabaticity would likely be destroyed by larger-scale fluid motions before significant helicity conversions have time to transpire. 

While we only sample discrete locations within the merger, what is a narrow pencil in resonant/adiabatic directions for a specific location will sweep out arcs when moving to nearby conditions. This could make it much more likely that a given neutrino will pass through an adiabatic helicity resonance somewhere, but the magnitude of this effect is difficult to quantify, and we consider it unlikely to make a significant number of neutrinos undergo helicity transformation.

In agreement with previous studies, we show that a large fraction of the volume is unstable to the neutrino fast flavor instability, and most of the locations exhibiting helicity resonance are unstable (Figure~\ref{fig:conditions_init}). We perform simulations of the fast flavor instability in a subset of the domain, and demonstrate that the instability modifies the strength and directional structure of the helicity-changing Hamiltonian (e.g., Figure~\ref{fig:hlr_time_evolution}). This results in a shift in the structure of the directional resonance bands (Figure~\ref{fig:resonant_directions_final}), and breaks degeneracies between helicity transformation modes (Figure~\ref{fig:linear_resonance_zoomed}). Although the fast flavor instability erases helicity resonance in some locations and creates resonance in others, the magnitude of the effect is not large enough to significantly increase the overall expected number of neutrinos that change their helicity state.

Although this study corroborates the claim that no significant helicity conversion occurs, we do not consider effects from magnetic fields on the neutrino magnetic moment, and do not consider implications for Dirac neutrinos. Our analysis is limited to a single snapshot of a single neutron star merger simulation, and different stages of the merger could lead to different regions experiencing helicity resonance (although the suppression from the neutrino mass would make a departure from these conclusions unlikely). Finally, the actual amount of helicity transformation for a given neutrino should ideally be computed with calculations similar to those carried out for simplified models (e.g., \cite{Tian2017}), but this will be left to future work.

\section{Acknowledgements}
HP was supported by the N3AS Physics Frontier Center through NSF award number 2020275. SR was supported by a National Science Foundation Astronomy and Astrophysics Postdoctoral Fellowship under award AST-2001760.
FF is supported by the Department of Energy, Office of Science, Office of Nuclear Physics, under contract number DE-AC02-05CH11231, by NASA through grant 80NSSC22K0719, and by the National Science Foundation through grant AST-2107932. { AVP would like to acknowledge support from the U.S. Department of Energy (DOE) under contract number DE-AC02-76SF00515 at SLAC National Accelerator Laboratory and DOE grant DE-FG02-87ER40328 at the University of Minnesota.} 

\bibliography{references}

\appendix

\section{Proof of Generalized Resonance Condition}
\label{app:proof}
Here we derive Equation \ref{eq:general_resonance}, a necessary condition for resonant helicity oscillations that can be applied in the multi-flavor case. 

First some notational definitions: given a state-vector $\ket{v}$ we define the subscript notation $\ket{v}_{\!L}$, $\ket{v}_{\!R}$ to represent the vector's unnormalized projections onto the left- and right-handed subspaces, respectively, with duals given by 
$\tensor[_L]{\bra{v}}{}$, $\tensor[_R]{\bra{v}}{}$. For example, a 2-spin state-vector $\ket{v} = \tfrac{1}{\sqrt{2}}(\ket{l} + \ket{r})$ would have $\ket{v}_{\!L} = \tfrac{1}{\sqrt{2}}\ket{l}$ (setting the right-handed part to 0). With this notation in mind we can begin the proof.

Consider a spin-1/2 neutrino state-vector $\ket{\nu(t)}$ of $N_F$ neutrino flavors that evolves under a Hamiltonian $H$ with eigenvectors $\ket{\lambda_j}$ and eigenvalues $E_j$; the time evolution of the state is then given by 
\begin{equation}
    \ket{\nu(t)} = \sum_{j=1}^{2N_F} a_j e^{-iE_j t} \ket{\lambda_j}
\end{equation}
for some coefficients $a_j = \braket{\lambda_j|\nu}$. 

 Assume that at $t=0$ the state is purely left-handed, and that the state exhibits resonant helicity oscillations so that at a later time $t'$ the state is purely right-handed. Using our projection notation we can write 
\begin{align}
   \left|\ket{\nu(0)}_{\!R}\right| = \left| \sum_{j=1}^{2N_F} a_j \ket{\lambda_j}_{\!R}\right| &= 0 \, , 
  \label{eq:resonant_ket_decomposition_0}\\
   \left |\ket{\nu(t')}_{\!R}\right | = \left | \sum_{j=1}^{2N_F} a_j e^{-iE_j t'} \ket{\lambda_j}_{\!R} \right| &= 1\, .
    \label{eq:resonant_ket_decomposition_1}
\end{align}
Without loss of generality, assume $\left |a_1 \right | \left | \ket{\lambda_1} \right | \geq \left |a_j \right | \left |  \ket{\lambda_j} \right |$ for all $j$. 

Taking the inner product with $\ket{\lambda_1}_{\!R}$ in Equation \ref{eq:resonant_ket_decomposition_0}, we get
\begin{align}
&\sum_{j=1}^{2N_F} a_j \tensor[_R]{\braket{\lambda_j|\lambda_1}}{_{\!R}} = 0 \\
\implies & -a_1 \tensor[_R]{\braket{\lambda_1|\lambda_1}}{_{\!R}} = \sum_{j=2}^{2N_F} a_j\tensor[_R]{\braket{\lambda_j|\lambda_1}}{_{\!R}}\\
\implies & \left |a_1 \right | \left | \ket{\lambda_1}{_{\!R}} \right |^2 \leq \sum_{j=2}^{2N_F} \left | a_j \right | \left | \tensor[_R]{\braket{\lambda_j|\lambda_1}}{_{\!R}} \right |
\label{eq:triangle_ineq_resproof}
\end{align}
where we have applied the triangle inequality in the last line. We will now manipulate Equation \ref{eq:triangle_ineq_resproof} to arrive at an inequality involving only $\left | \tensor[_R]{\braket{\lambda_1|\lambda_1}}{_{\!R}} \right |$, which will give us a restrictive condition on resonance.

We first use the Cauchy-Schwartz inequality on the RHS of Equation \ref{eq:triangle_ineq_resproof}, treating the ${\left | a_j \right | }$ and $\left | \tensor[_R]{\braket{\lambda_j|\lambda_1}}{_{\!R}} \right |$ as components of $2n-1$ dimensional vectors:
\begin{equation}
    \sum_{j=2}^{2N_F} \left | a_j \right | \left | \tensor[_R]{\braket{\lambda_j|\lambda_1}}{_{\!R}} \right |
    \leq \sqrt {\sum_{j=2}^{2N_F} \left | a_j \right | ^2} \sqrt {\sum_{j=2}^{2N_F} \left | \tensor[_R]{\braket{\lambda_j|\lambda_1}}{_{\!R}} \right |^2} \, .
    \label{eq:cauchy_schwartz_resproof}
\end{equation}

 The first root on the RHS of \ref{eq:cauchy_schwartz_resproof} can be simplified via the normalization condition of the initial state: 
 \begin{equation}
 \sum_{j=2}^{2N_F} \left|a_j\right |^2 = 1 - \left|a_1\right|^2 \, .
 \label{eq:solving_root1_resproof}
 \end{equation}

 To simplify the other root, first note that $\braket{ \lambda_j|\lambda_1}_{\!R} = \tensor[_R]{\braket{\lambda_j|\lambda_1}}{_{\!R}} + \tensor[_L]{\braket{\lambda_j|\lambda_1}}{_{\!R}} = \tensor[_R]{\braket{\lambda_j|\lambda_1}}{_{\!R}},$ and hence that
\begin{align}
    &\sum_{j=2}^{2N_F} \left | \tensor[_R]{\braket{\lambda_j|\lambda_1}}{_{\!R}} \right |^2 = \sum_{j=2}^{2N_F} \left | \braket{ \lambda_j|\lambda_1}_{\!R} \right |^2 \\
    =& \sum_{j=1}^{2N_F} \left | \braket{ \lambda_j|\lambda_1}_{\!R} \right |^2 - \left | \braket{ \lambda_1|\lambda_1}_{\!R} \right |^2  \\
    =& \left | \ket{\lambda_1}_{\!R} \right |^2- \left | \ket{\lambda_1}_{\!R} \right |^4 \, . 
    \label{eq:solving_root2_resproof}
\end{align}

Substituting \ref{eq:solving_root1_resproof} and \ref{eq:solving_root2_resproof} into the RHS of Equation \ref{eq:cauchy_schwartz_resproof}, combining with Equation \ref{eq:triangle_ineq_resproof}, and rearranging, we get 
\begin{equation}
    \frac{|a_1|}{\sqrt{1 - \left|a_1\right|^2}} \leq
       \frac{\sqrt{1- \left | \ket{\lambda_1}_{\!R} \right |^2}}{\left | \ket{\lambda_1}_{\!R} \right |} \, .
       \label{eq:main_inequality_resproof}
\end{equation}
We now seek to eliminate $|a_1|$ from the above inequality to get an expression involving only $\left | \ket{\lambda_1}_{\!R} \right |$.

Recalling Equation \ref{eq:resonant_ket_decomposition_1}, we have that
\begin{align}
    1 =&  \left | \sum_{j=1}^{2N_F} a_j e^{-iE_j t'}\ket{\lambda_j}_{\!R} \right| \\  
    \leq &\sum_{j=1}^{2N_F} \left | a_j e^{-iE_j t'} \ket{\lambda_j}_{\!R} \right| =\sum_{j=1}^{2N_F} \left | a_j \right| \left |  \ket{\lambda_j}_{\!R} \right| \\
    \leq & \sum_{j=1}^{2N_F} \left | a_1 \right| \left | \ket{\lambda_1}_{\!R} \right| = 2N_F\left | a_1 \right| \left |\ket{\lambda_1}_{\!R} \right| \, ,
\end{align}
using the triangle inequality and the fact that $\left |a_1 \right| \left |\ket{\lambda_1} \right | \geq \left |a_j \right| \left |\ket{\lambda_j} \right |$ for all $j$. It follows that
\begin{equation}
\left | a_1 \right | \geq  \frac{1}{2N_F\left |\ket{\lambda_1}_{\!R} \right|} \geq (\frac{1}{2N_F})
\end{equation}
since $\left |\ket{\lambda_1}_{\!R} \right| \leq 1$, and thus 
\begin{equation}
\frac{|a_1|}{\sqrt{1 - \left|a_1\right|^2}} \leq 
\frac{(\frac{1}{2N_F})}{\sqrt{1 - (\frac{1}{2N_F})^2}}
\end{equation}
since this is an increasing function on $0 \leq|a_1| \leq 1.$ Inserting into Equation \ref{eq:main_inequality_resproof}, we have 
\begin{equation}
   \frac{(\frac{1}{2N_F})}{\sqrt{1 - (\frac{1}{2N_F})^2}} \leq
      \frac{\sqrt{1- \left | \ket{\lambda_1}_{\!R} \right |^2}}{\left | \ket{\lambda_1}_{\!R} \right |}  \, .
\end{equation}
Solving for $\left | \ket{\lambda_1}_{\!R} \right |$ gives 
\begin{equation}
    \left | \ket{\lambda_1}_{\!R} \right |^2 \leq 1 - (\tfrac{1}{2N_F})^2 \, ,
\end{equation}

and since $\left | \ket{\lambda_1}_{\!R} \right |^2 + \left | \ket{\lambda_1}_{\!L} \right |^2 = 1$ (as this is an orthogonal decomposition of a unit vector) we can also infer that 
\begin{equation}
    \left | \ket{\lambda_1}_{\!L} \right |^2 \geq \left (\frac{1}{2N_F} \right)^2 \, .
\end{equation}
Since our argument applies equivalently if we invert the $L-$ and $R-$subspaces, we can equivalently infer that resonance requires 
\begin{align}
    \left | \ket{\lambda_1}_{\!R} \right |^2 &\geq \left( \frac{1}{2N_F} \right)^2 \, ,\\
    \left | \ket{\lambda_1}_{\!L} \right |^2 &\leq 1 - \left(\frac{1}{2N_F}\right)^2 \, .
\end{align}
 
 Combining these inequalities, we have that a necessary condition for there to exist some state presenting resonance is that at least one eigenvector $\left | \ket{\lambda_1} \right |$ of the Hamiltonian satisfies
\begin{equation}
    \left | \vphantom{\sum}  \left | \ket{\lambda_1}_{\!R} \right |^2 - \left | \ket{\lambda_1}_{\!L} \right |^2 \,  \right | \leq 1 - 2\left(\frac{1}{2N_F}\right)^2 
\end{equation}

or as expressed in this paper, the generalized resonance condition is expressed in Equation~\ref{eq:general_resonance}.
\end{document}